\documentclass[12pt]{amsart}
\usepackage{geometry}                
\usepackage{graphicx}
\usepackage{setspace}
\usepackage{amssymb}
\usepackage{epstopdf}
\usepackage{setspace}
\graphicspath{ {images/} }
\title{Geometry and Physics of $Sp(3)/Sp(1)^3$}
\author{B. E. Eichinger}
\address{Department of Chemistry, University of Washington, Seattle, WA 98118}
\begin{document}
\maketitle
\section*{Abstract}
The action of $Sp(3)$ on a vector space $V_3\in \mathbb H^3$ is analyzed.  The transitive action of the group is conveyed by the flag manifold (coset space) $Sp(3)/Sp(1)^3\sim G/H$, a Wallach space. The curvature two-forms are shown to mediate pair-wise interactions between the components of the $\mathbb H^3$ vector space.  The root space of the flag manifold is shown to be isomorphic to that of $SU(3)$, suggesting similarities between the representations of the flag manifold and those of $SU(3)$.    The passage from $SU(3)$ to $Sp(3)$ and the interpretation given here encompasses the spin of the fermionic components of $V_3$.  Composite fermions are representable as linear combinations of product states of the eigenvectors of $G/H$. 

\section*{Introduction: Flags and Flag Manifolds}  

Matter is built up from elementary particles: quarks comprise nucleons, nuclei and electrons make up atoms, atoms bond together to form molecules, which might comprise a crystal that is placed in an instrument that is located in a laboratory in a building $\cdots$.  The sequence can be continued to encompass as much of the world as one likes.  This statement can be rendered abstractly: there exists a sequence of subsets of material objects, $S_1\subset S_2\subset  S_3\subset  \cdots$, where $S_1=$quarks, $S_2=$nucleons, $S_3=$nucleons + electrons = atoms,  $S_4=$ molecules, \emph {etc.}, of some larger set of objects.  Since all matter is composed of elementary particles, each subset in the sequence is built up from the members of the preceding subset.  There is also an implied geometry of spatial inclusion as one climbs the ladder of complexity.  By imposing a (perhaps abstract) geometrical relation between objects we can do more than talk about sets and subsets; there is a mathematical structure that accommodates these notions.  A \emph{flag} is a sequence of vector subspaces: $V_1\subset V_2\subset V_3\subset\cdots\subset V_n$ of a vector space $V_n$ of dimension $n$ determined by the largest space of interest, equivalent to truncating our sequence of objects at some desired level.  These elementary observations motivate an interest in flags and their associated flag manifolds. 

In the standard definition, a flag is a sequence of vector spaces $V_i(\mathbb{K})$ over the real or complex field $\mathbb {K=R,C}$ such that $V_i$ is a proper subset of $V_{i+1}$ for all $i$ up to the complete space $V_n$. (See Wikipedia for an introduction.  As a further aside, a search of \emph{arXiv} reveals a flurry of recent work on flag manifolds.) The definition of the flag extends to the quaternion ring $\mathbb {K=H}$, which will be central to the theory presented here.  There is a natural action of a Lie group $H\subset G$ on the flag that preserves the flag structure, while the Lie group $G$ acts on the entire flag.  The flag manifold is constructed from cosets $G/H$ of the group in a manner to be discussed.

$SU(3)$ can be interpreted as a group acting on an abstract vector space $V_3(\mathbb C)$ of quarks. Within the group there are no space-time coordinates, which implies that the group is describing an underlying abstract geometry or symmetry of the three-component object.  This symmetry is an intrinsic property, defined by the group without reference to other objects.  Thus all protons, for example, have identical intrinsic properties, independent of their location.  The flag manifold $SU(3)/U(1)\times U(1)\sim U(3)/U(1)\times U(1)\times U(1)$ associated to the unitary group has been investigated recently.\cite{Byk,Am} The non-linear sigma model over higher dimensional $\mathbb C^n$-flags has also been developed.\cite{Sei}

Our aim is to transition from $SU(3)$ to $Sp(3)$ so as to directly incorporate spin degrees of freedom. 
The presentation begins with a discussion of a few of the algebraic and analytical aspects of the pertinent groups that are essential to understand the overall structure of the flag and flag manifold.  A practice calculation with $Sp(2)/Sp(1)^2$, related to Yang-Mills theory, will set the stage for the main results.  As the development progresses, several general mathematical structures applying more widely to flags than the $\mathbb H^3$ case will emerge.  The restricted goal, which will occupy most of our attention, is to uncover the relation between representations of $SU(3)$ and those of the $Sp(3)/Sp(1)^3$ Wallach space.\cite{Wal} As the geometry develops, it will be shown that curvature operators act on the $V_3(\mathbb H)$ components, thereby yielding a representation of the forces acting between the elementary particles.

\subsection*{Structural Preliminaries}The mathematical structure introduced above has been formulated as a theory of interactions.\cite{BEE1}  A group represents the interactions while the module on which the group acts is a state space of a many-body system, similar to that concept in statistical mechanics. By choosing the group to preserve a measure on the total space $V_n$, the group is required to be compact.  A reducible representation of the group corresponds to two or more disjoint spaces, so it is natural to think about systems that interact with one another to correspond to irreducible representations (irreps).  Systems that are deemed to be independent of one another, or that do not measurably perturb one another, may be treated independently -- their representation spaces are \emph{effectively} orthogonal.  This enables one to truncate a very large flag at any point that is deemed to be an acceptable representation of an isolated system.  There are many additional consequences of these statements, some of which are related to interpretations of quantum mechanics, which will be elaborated elsewhere.  

The general setting for the theory is provided by a principal bundle,\cite{KN} generically written as $G(G/H,H)$, where $G$ is the bundle space, $G/H$ is the base space, and $H$ is the fiber.  In the flag context, $G/H$ is the flag manifold, and $H$ fixes the components of the flag.  Our primary focus will be on  $H=\bigotimes^n_{i=1} h_i(\mathbb K)$, which acts on $x\in V_n$ by 
\[
HV_n:=\textrm{diag}[h_i]\left[{\begin{array}{*{20}c}
v_1\\
v_2\\
.\\
.\\
v_n
\end{array}}\right]=\left[{\begin{array}{*{20}c}
h_1v_1\\
h_2v_2\\
.\\
.\\
h_nv_n
\end{array}}\right]=\hat V_n,
\]
This subgroup leaves the components of the flag point-wise fixed in the vector space $V_n(\mathbb K)$.  (More general flags will be considered later.)  That is, only the $x\in G/H$ component of $g=xh: g\in G,h\in H$,  changes the magnitudes of the individual $v_i$, i.e., acts transitively on $V_n$.  The gauge group $h_i$ is (\emph{i}) the trivial identity for $\mathbb {K=R}$, (\emph{ii})supplies a phase change or rotation for $\mathbb {K=C}$, and (\emph{iii}) acts as a rotation of a basis vector for $\mathbb {K=H}$.  The action of $H$ establishes an effective isomorphism between the diagonal elements of the group and the basis vectors of $V_n$.  Note that $H$, as well as $G$, leaves the bilinear form $\langle W,V\rangle$ invariant.  The vector space also accommodates a right action, $(g,\bar h)V_n\to gv\bar h$, where $\bar h$ is compatible with $v_i$ on the right.  This is a ``global phase change", but this extension will not be further considered here.
 
In $SU(3)$, higher dimensional representations than the fundamental are represented as linear combinations of the fundamental root vectors, and these vectors organize the symmetries of mesons and baryons.  In a general setting, higher dimensional irreps might encompass excited states, and a group can be imagined to move excitations from one component to another.  This interpretation identifies elements of the flag manifold with bosons, while the vector space (or module) on which the group acts consists of fermions, as averred in the Abstract.  In addition, higher dimensional representations enable analytic expressions for composite states to be built from combinations of the elementary units, and the quaternion algebra allows this to be done while preserving the spin 1/2 structure. 

The spin (or isospin) of an elementary particle is incorporated at the outset in this description by choosing $\mathbb {K=H}$. Since we want a group structure, quaternions rather than Pauli matrices are used.  The motivation for selection of $\mathbb {K=H}$ is that, of the several properties of fundamental particles -- mass, charge, and spin -- only the last is intrinsically based on a group action.  These ideas direct attention to the symplectic group $Sp(n)$, which is a compact topological space over $\mathbb H$.  In the physical context the group acts on a Hilbert space, which is interpreted as the state space of an $n$-body system assembled into several parts conforming to a flag description.  In application to fundamental particle systems, the cosets (complete flag manifolds) $Sp(k)/Sp(1)^k:=Sp(k)/Sp(1)\times\cdots\times Sp(1)$ for small $k$ are of interest.  The group $Sp(n)$ consists of matrices that are unitary over the quaternions: $Sp(n)\sim U(n,\mathbb H)$.  $Sp(n)$ is a compact subgroup of $Sp(2n,\mathbb C)$; its Lie algebra $\mathfrak{sp}(n)\subset\mathfrak{sp}(2n,\mathbb C)$ is contained in that of the larger group.  (Sources for group theory are refs. \cite{Helg,FH,Simon,GW}.)

An extremely important aspect of a group action is that geodesics on a Lie group are left cosets of one parameter subgroups.\cite {Simon, Price, Sternberg}  A map from $t\in \mathbb R$ to a coset is constructed by first selecting an $\mathfrak {x\in g\backslash h}$, \emph {i.e.} ${\mathfrak x\notin\mathfrak h}$, where $\mathfrak {g}$ is the Lie algebra of the group and $\mathfrak h$ is the algebra of the fiber.  The map $\exp: t\to \exp(t\mathfrak x), t\ge 0$, is a geodesic through the origin in the base space $G/H$, which introduces a \emph{global} time coordinate.

The third important structure involves the separation of the action of $H$ and $G/H$ on the Hilbert space of higher dimensional representations.\cite{Folland}  Define a function $\Psi_\phi(x)$ in the representation space of $G$ by 
\[
\Psi_\phi(x):=\int\sigma(h)\phi(xh)dh
\]
where $dh$ is normalized Haar measure on the group $H$, $x\in G/H, h\in H$ and $\phi(xh)$ is a map from $G$ into a Hilbert space of dimension compatible with the representation $\sigma(h)$; $\sigma(ab)=\sigma(a)\sigma(b)$ for $a,b\in H$; $\sigma$ is a homomorphism.  Given a left invariant Haar measure on $H$ it follows that 
\begin{equation}\label{ind}
\Psi_\phi(x\eta):=\int\sigma(h)\phi(x\eta h)dh=\sigma(\eta^{-1})\Psi_\phi(x)
\end{equation}
for $\eta\in H$.  The $L^2$ measure $\langle\Psi_\phi(x),\Psi_\phi(x)\rangle$ on the Hilbert space is thus invariant to $\sigma(h)$.  For $H=H_1\times H_2$ there is an isomorphism $h_1\times h_2\to \sigma(h_1)\times \sigma(h_2)$.  By averaging a representation over the fiber, one obtains an left action of $\sigma(H)$, valid for representations of any dimension, thereby mimicing the action of $H$ on $V_n$ in the fundamental representation.  This enables us to mix together elementary reps of $V_n$ with those of composites, which will prove useful.  Eq. (\ref{ind}) makes explicit the dependence of reps $\Psi_\phi(x)$ on the flag manifold, independent of the gauge group, and also suggests that one might construct $\Psi_\phi(x)$ directly.

Under the action of $g\in G$, the coset $xH$ is sent to $g:xH\to yH$.  A representation $A_g$ acts on the left by $A_g\Psi(x)=\Psi(g^{-1}x)=\Psi(yh)=\sigma(h^{-1})\Psi(y)$. (See the parallel presentation for finite groups in Sec. V.3 of ref. \cite{Simon}.)

\subsection*{Interlude:  Generalizations and Yang-Mills Theory}
The flag manifold structure can be extended to any $U(n)=U(n,\mathbb K),n>2$, and partition of $n$:
\[
\{k_1,k_2,\cdots,k_m\};\; \sum^m_ik_i=n, 
\]
with corresponding coset $U(n)/U(k_1)\times U(k_2)\times\cdots\times U(k_m)$, consistent with the subspace decomposition, $V_1\subset V_2\subset V_3\subset\cdots\subset V_n$; $k_1=\textrm{dim}(V_1), k_i=\textrm{dim}(V_i)-\textrm{dim}(V_{i-1}),2\le i\le n$ .  Many recent papers  \cite{Byk,Am,Sei} have explored this generalization with  $\mathbb{K=C}$ in the context of a (generalized) nonlinear sigma model.  Our focus is not a sigma model, but is a different, direct approach to understand natural, composite spin 1/2 particles.  The confluence of interests in flag manifolds may be coincidental, or there may be connections that are unforeseen.  One advantage of the direct approach here is that it yields metrics, Lie algebra operators, and curvature two-forms or tensors that are directly related to forces.    

In anticipation of these results, it can now be recalled that a crucial example of a structure conforming to all of the above has long been known.  Atiyah showed that the Yang-Mills functional\cite{YM,BPST, Atiyah} is minimized by the curvature two-form (his notation)
\[
F=\frac{dq\wedge d\bar q}{(1+q\bar q)^2},
\]
where $\bar q$ is the conjugate of the quaternion $q$.  The geometrical structure that yields this curvature two-form is the Grassmannian $Sp(2)/Sp(1)^2$.\cite {Atiyah, Lawson} One may interpret this coset space structure in the fundamental representation as the action of $Sp(2)$  on a square-integrable, quaternion-valued Hilbert space $V_2(\mathbb H)$, with the subgroup $Sp(1)\times Sp(1)$ acting on each $V_1(\mathbb H)$ separately, as was discussed above.  The well-known Lie algebra isomorphism $\mathfrak{sp}(1)\sim\mathfrak{su}(2)\sim\mathfrak{so}(3)$ then leads to the identification of the two $Sp(1)$ components of the group $Sp(1)\times Sp(1)$ of the fiber as spin (or gauge groups of isospin in the Yang-Mills context) degrees of freedom of single particle states.  The coset, $Sp(2)/Sp(1)\times Sp(1)\sim q$ consists of the instanton coordinates, and the action of the coset can be interpreted as a coupling of the two components of $V_2(\mathbb H)$; the two elementary or fundamental quaternions affect one another through the action of the coset.  The curvature of the coset space is equivalent to an interaction or force between the particles, as conveyed by the curvature form above.  The separation of the components of the group that is given by eq. (\ref{ind}) shows how the representation space of a principal bundle relates the gauge group of isotopic spin to the instanton content of the representation space.  

Given  the flag and flag manifold structure, and the encouraging confirmation from Yang-Mills theory, the next natural extension is to a system of three fundamental fermions, which should be described with $Sp(3)/Sp(1)^3$.  Knowing that $SU(3)$ is a subgroup of $Sp(3)$ is sufficient motivation to ask whether the larger group and coset might provide additional insight into the structure of mesons and baryons.  Indeed it does; we will be able to construct explicit functions to describe physical states from the eigenspace of the Lie algebra of the coset. 

The scope of the present work is limited to showing how composite particles might be constructed from the eigenspaces of $Sp(n)/Sp(1)^n$ for $n=\{2,3\}$.  Most importantly, it will be shown that the coset space $Sp(3)/Sp(1)^3$ has a rank two algebra, and the corresponding root space is surprisingly isomorphic to that of $SU(3)$.  The primary objective is tool development, and while a few comments about representations will be offered, detailed assignments are not considered.  Perhaps experts will find the preliminary assignments sufficiently interesting to encourage their participation.  

In the following, \emph {spin} will simply mean the quaternion content of the structures to be developed; the reader may prefer to use \emph{isotopic spin} and the $Sp(1)$ fibers as \emph{gauge groups}.  (However, our interpretation of these concepts may differ in some respects.)  The fact that there is an instanton in the theory means that this is not a relativistic theory.  The departure from relativity also relates back to the use of quaternions rather than Pauli matrices to describe spin.  There are mappings between compact and hyperbolic spaces when restricted to a single interaction,\cite{BEE2} so connections with relativity can be made.  

\subsection*{Algebraic Preliminaries}
A quaternion, $q$, is represented in the Hamiltonian basis as $q=q_0{\bf 1}+q_1{\bf i}+q_2{\bf j}+q_3{\bf k}: q_i\in \mathbb R, 0\le i\le 3$, where the anti-commuting basis elements of the algebra, $\{{\bf 1,i,j,k}\}$, satisfy ${\bf ii=jj=kk= ijk=-1}$.  The quaternion algebra is associative and distributive, but not commutative (except for the unit element {\bf 1}).  Hamilton's mixed scalar-vector notation, $q=q_0{\bf 1}+{\bf q}$, where ${\bf q}$ is the vector/imaginary part of the quaternion, is often handy for calculations. The product of two quaternions, $a$ and $b$, is $ab=(a_0b_0-{\bf a\cdot b}){\bf 1}+a_0{\bf b}+b_0{\bf a}+{\bf a\times b}$.  The advantage of the mixed notation is that standard vector operations, scalar and vector product as used here, is useful shorthand. (There are hazards in the use of the scalar-vector notation.  Writing the vector part as a quaternion product, ${\bf rr=-r\cdot r+r\times r=-|r|^2}$; note that the scalar product contains a center dot: ${\bf r\cdot r=+1}$.)  The conjugate quaternion is $\bar q=q_0-{\bf q}$.  Using the product rule it is easy to show that $q\bar q=|q|^2=q^2_0+q^2_1+q^2_2+q^2_3$, where $|q|$ is the norm of the quaternion.  

In many calculations to follow the norm of a quaternion factors from the problem or is otherwise of secondary importance, so that only the unit part, $u$, of $q=|q|u$ is of interest.  The logarithm of a quaternion exits in the sense that a unit quaternion has an exponential form: $u=\exp(v)$.    Since $u\bar u=\bar uu={\bf 1}$, it follows that $\bar u=u^{-1}\Longrightarrow \bar v=-v$, signifying that $v={\bf v}$ is a purely imaginary quaternion.  (A unit quaternion is isomorphic to the three sphere $S^3$, and the tangent space of $S^3$ is isomorphic to ${\mathbb R^3}$.)  Expanding the exponential,
\[
\exp({\bf v})={\bf 1}+{\bf v}-\frac{1}{2!}|v|^2{\bf 1}-\frac{1}{3!}|v|^2{\bf v}+\frac{1}{4!}|v|^4{\bf 1}+\cdots=\cos(|v|){\bf 1}+|v|^{-1}\sin(|v|){\bf v},
\]
which makes clear that the magnitude, $|v|$, of ${\bf v}=|v|{\bf r}$, also factors, so that a quaternion $q$ may be expressed as $q=|q|\exp(\chi{\bf r})$, where $\chi=|v|$.  While there is no natural restriction on $\chi$ when a quaternion acts as an operator, functions $\Psi(u)$ that appear in a physical context may require periodicity conditions.   It is also clear that $\bar u=u^{-1}=\cos(\chi){\bf 1}-\sin(\chi){\bf r}$.  In the $Sp(n)$ context, the representation $q=|q|\exp({\bf v})$ recommends against identifying the component $q_0{\bf 1}$ with a temporal variable, but there is more to this story that will emerge as the theory develops.

The Hamiltonian basis might also be described as the $Sp(1)$-basis or $\mathbb H$ representation.  There is also a well-known $SU(2)$ basis or $\mathbb C^2$-basis (and an inclusion $\mathfrak{sp}(n)\subset\mathfrak{sp}(2n,\mathbb C)$ of  Lie algebras), which enables one to identify isomorphic basis elements in the usual way:
\[
{\bf 1}\sim\left[{\begin{array}{*{20}c}
1& 0\\
0 & 1\\
\end{array}}\right],\quad {\bf i}\sim\left[{\begin{array}{*{20}c}
0& 1\\
-1 & 0\\
\end{array}}\right],\quad{\bf j}\sim\left[{\begin{array}{*{20}c}
0& i\\
i & 0\\
\end{array}}\right],\quad{\bf k}\sim\left[{\begin{array}{*{20}c}
i& 0\\
0 & -i\\
\end{array}}\right]; \quad i=\sqrt{-1}
\]
which gives a conventional form for a quaternion as 
\begin{equation}\label{conform}
q=\left[{\begin{array}{*{20}c}
q_0+iq_3& q_1+iq_2\\
-q_1+iq_2& q_0-iq_3\end{array}}\right]=\left[{\begin{array}{*{20}c}
\zeta_1& \zeta_2\\
-\bar \zeta_2 &\bar\zeta_1 \end{array}}\right].
\end{equation}
There are specific advantages to the use of both representations.  

Many constructions to be encountered involve the trace operation over a product of quaternion matrices.  In general, $\textrm{tr}(AB)\ne \textrm{tr}(BA)$ for $\{A,B\}$ compatible matrices over $\mathbb H$.  However, if the diagonal elements of the $AB$ product are pure real, the cyclic permutation rule is valid.

Within the $SU(2)$ basis there is an operation that is extremely useful for computations.  Define $J\sim{\bf i}$; a small calculation shows that \emph{complex conjugation}: $a\to \bar a$ in this basis is accomplished with 
\begin{equation}\label{Jop}
\bar a=J'aJ=-JaJ=JaJ';
\end{equation}
this is just a rotation by $\pi$ around the ${\bf i}$-axis (see below).  The \emph{quaternion conjugate} in the matrix basis is $a^*=\bar a'=J'a'J$, where $a'$ is the transpose of $a$.  Use of the $J$-operator facilitates computation of derivatives in the $SU(2)$ basis, while Hamilton's scalar-vector notation is useful for algebraic calculations. (The use of $\bar a$ to signify conjugation for both the $Sp(1)$ and $SU(2)$ representations has to be handled with care, as they are not equivalent.  The equivalence is $\bar a_{\mathbb {H}}\sim a^*_{\mathbb {C}^2}$; the context should make it clear which is intended.  In the multi-dimensional case, say a matrix $A\to A^*$, both conjugation and transposition are intended, so the meaning of $A^*$ is unequivocal.)

There are three involution operations on quaternions that may be equivalent to the CPT operators of quantum theory.  Working in the $\mathbb H$-basis, parity is clearly $P:a\to \bar a$, as seen above.  The other two are reversal of the identity component, $I:a\to -\bar a$, and simple negation, $N:a\to -a$.  $PIN$ in any order is the identity when operating on a simple quaternion.  However, the operators are more interesting when acting on products, $ab$.  Now $P:ab\to \bar b \bar a\ne P(a)P(b)=\bar a\bar b$.  Similarly, $I:ab\to -\bar b\bar a\ne I(a)I(b)=(-\bar a)(-\bar b)=\bar a\bar b$, and $N:ab=-ab\ne N(a)N(b)=(-a)(-b)=ab$, yet each squares to the identity, as one can easily prove.  In addition to $PIN$, the operators ${\bf e}_m=\{\bf{i,j,k}\}$ acting by ${\bf \bar e}_m a{\bf e}_m=-{\bf e}_m a{\bf e}_m$ (one of which we've seen acting in the $SU(2)$ representation as complex conjugation) is a rotation by $\pi$ about the ${\bf e}_m$ axis.  (For evaluating $PIN$ in the $SU(2)$-basis, make the substitution $\bar x\to x^*$.)

The quantum mechanical parity operator, $P:{\bf x}\to -{\bf x}, {\bf x}\in \mathbb R^3$, is equivalent to conjugation: $u\to \bar u$.  The left, or right, action of the quaternion $\exp(-2{\bf v})$ on $u=\exp({\bf v})$ gives the conjugate.  However, care must be taken to distinguish this algebraic operator from the abstract parity operator $P$, for which $P(Pu)=P(\bar u)=u \Longrightarrow P^2=1$.  Clearly $\exp(-2{\bf v})$ does not square to the identity.  The point is that conjugation can be realized by the multiplicative action of an appropriate quaternion residing in $Sp(n)$.  This provides explicit operators that execute transitions from a state with positive chirality to one of negative chirality and \emph{vice-versa}.  Since there is no notion of a direction of motion in this discussion, the word ``helicity" is avoided.

\section*{General Structure of the Eigenvalue Problem on $Sp(n)/Sp(1)^n$}

The representations of the classical groups are well known.\cite{FH}  However, representations parameterized by cosets are apparently less well documented.  Given eq. (\ref{ind}), and the desire to construct explicit representations to show how composite states are realized, we will use a direct, na\"ive approach to calculate eigenvalues and eigenvectors.  This will show how the matrix elements of the coset are related to the module on which the group acts, and it is anticipated that this will provide some insight into the relation between bosons and fermions.  

As shown above, the parameterization of $Sp(n)$ that we will be working with is built on the coset $xH$ structure, such that an element $g\in Sp(n)$ is written as $g=xh$, where $x=\exp(\mathfrak x)$ and $h\in H$ is a diagonal matrix, all elements of which are unit quaternions.  Since $gg^{-1}=gg^*=xhh^*x^*=xx^*=1$, it follows that the Lie algebra $\mathfrak x$ of $x$ is skew-symmetric: $\mathfrak x^*=-\mathfrak x$.  The diagonal elements of $\mathfrak x$ are identically zero, as they have been pulled into $\mathfrak h$, where $\exp(\mathfrak h)=H$.  It is useful to introduce some notation.  The fundamental representations of $Sp(n)/Sp(1)^n$ do not  represent the whole group; clearly the maximal subgroup $H$ is excluded from the representation, and this implies that the rank of the "root" space of the coset is less than that of the whole group.  Let $CF(n,\mathbb H)$ denote the subgroup of $Sp(n)$ that is parameterized by the components of the coset, $Sp(n)/Sp(1)^n$, where $CF$ suggests Complete Flag.  Since we are working exclusively in $\mathbb H$ or the isomorphic $\mathbb {R}^+\times SU(2)$ presentations, this will be simply $CF(n)$.

The eigenvalue problems to be solved for $x\in Sp(n)/Sp(1)^n$ are $\Lambda=\tau^*x\tau=\tau^*(\exp{\mathfrak x})\tau=\exp(\tau^*{\mathfrak x}\tau)=\exp(\lambda)$ since $\tau^*\tau=\tau\tau^*=1$. Here $\{\Lambda,\lambda\}$ are diagonal matrices in either the $\mathbb H$ or $\mathbb C^2$ basis.  One solves for eigenvalues in the algebra rather than the group.  On selecting an $\mathfrak {x\in g\backslash h}$, geodesics are of the form $\exp(t\mathfrak x)$, so that the maximal torus of the group is $\exp(t\lambda)$.   These elementary statements about Lie groups are well known.  Also well known is that higher dimensional irreducible representations are constructed from tensor products
\begin{equation*}
\bigotimes_1^m x=x\otimes x\otimes\cdots\otimes x.
\end{equation*}
and that linear combinations of the elements of these products are classified by their symmetries with respect to interchanges of matrix elements.  By relating the eigenvectors to the matrix elements, we will uncover symmetry relations between bosons and fermions. 

\section*{Case 1: $Sp(2)/Sp(1)^2$}
\subsection*{Metric and Curvature}
The metric and curvature two-forms for the general Grassmannian $Sp(k+n)/Sp(k)\times Sp(n)$ will be presented here, even though the case $k+n=2$ is algebraically simpler.  The restriction to $Sp(2)/Sp(1)^2$ will be presented at the end of this section.  

A general matrix $g\in Sp(k+n), gg^*=g^*g=1$ that is partitioned to be compatible with the subgroup $Sp(k)\times Sp(n)$ is
\[
g=\left[{\begin{array}{cc}
A&B \\
C &D\\
\end{array}}\right]=\left[{\begin{array}{cc}
1&X\\
-X^* &1 \\
\end{array}}\right]\left[{\begin{array}{cc}
A&0 \\
0 &D\\
\end{array}}\right]
\]
where the $k\times n$ matrix $X=BD^{-1}=-(A^*)^{-1}C^*$.  The latter equality comes from the orthogonality $g^*g=1$.  Here $X^*$ is the transpose conjugate of $X$, a notation that covers both the $\mathbb H$ and $\mathbb C^2$ bases as noted above.  Orthogonality also yields $1+XX^*=(AA^*)^{-1}$ and $1+X^*X=(DD^*)^{-1}$.  The eigenvalues of $A$ and $D$ are determined by the eigenvalues of $X$.  The invariance of $AA^*$ and $DD^*$ to the right action of $h\in Sp(k)\times Sp(n)$:
\[
h= \left[{\begin{array}{cc}
h_k&0 \\
0 &h_n \\
\end{array}}\right]
\]
on $\textrm{diag}(A,D)$ provides an explicit representation of the coset structure $G/H$, i.e., $A= (1+XX^*)^{-1/2}h_k$ and $D=(1+X^*X)^{-1/2}h_n$, which is consistent with a count of real variables (a polar decomposition of $A$ and $D$ is implied here). 

The action of $g_1\in Sp(k+n)$ on $X$ is $g_1:X\to (A_1X+B_1)(C_1X+D_1)^{-1}$.  From these relations one may construct\cite{Hua, BEE1} the invariant metric on the coset space
\[
ds^2=\textrm{tr}[(1+XX^*)^{-1}dX(1+X^*X)^{-1}dX^*].
\]
Making use of $dX=dBD^{-1}-BD^{-1}dDD^{-1}=(A^*)^{-1}(A^*dB+C^*dD)D^{-1}$ (note that the subscript on the blocks of $g_1$ have been dropped), the metric can also be written 
\[
ds^2=\textrm{tr}[(A^*dB+C^*dD)(dB^*A+dD^*C)].
\]
The metric on $Sp(k+n)$ is $\textrm{tr}(dgdg^*)=-\textrm{tr}(g^*dgg^*dg)=\textrm{tr}(\omega\omega^*)$, where 
\[
\omega=g^*dg=\left[{\begin{array}{cc}
\omega_{11}&\omega_{12 }\\
\omega_{21}&\omega_{22 }\\
\end{array}}\right]=\left[{\begin{array}{cc}
\omega_{11}&\omega_{12 }\\
-\omega^*_{12}&\omega_{22 }\\
\end{array}}\right].
\]
is skew-symmetric since $d(g^*g)=0$.  The metric on the coset space is simply the trace of the square of the off-diagonal block of $\omega$, i.e., $ds^2=\textrm{tr}(\omega_{12}\omega^*_{12})$.

The left invariant differential form $\omega$ has an exterior derivative\cite{Cartan,Chern95}
\[
d\omega=dg^*\wedge dg=-g^*dg\wedge g^*dg=-\omega\wedge \omega
\]
which is the second Maurer-Cartan equation: $d\omega+\omega\wedge \omega=0$.  Writing this out in block form gives 
\[
d\omega+\omega\wedge \omega= \left[{\begin{array}{cc}
d\omega_{11}+\omega_{11}\wedge\omega_{11}+\omega_{12}\wedge\omega_{21}&d\omega_{12}+\omega_{11}\wedge\omega_{12}+\omega_{12}\wedge\omega_{22} \\
d\omega_{21}+\omega_{21}\wedge\omega_{11}+\omega_{22}\wedge\omega_{21}&d\omega_{22}+\omega_{22}\wedge\omega_{22}+\omega_{21}\wedge\omega_{12}\end{array}}\right]=0
\]
One can apply Cartan's criterion: $d\omega_{\mu\mu}+\omega_{\mu\mu}\wedge\omega_{\mu\mu}=\Omega_{\mu\mu}\sim \Omega_\mu$ for the diagonal elements to define (see Sec. 7 of \cite{Chern46} for a proof)  the curvature two-forms as
\begin{align*}
&\Omega_1=-\omega_{12}\wedge\omega_{21}=\omega_{12}\wedge\omega^*_{12}\\
&\Omega_2=-\omega_{21}\wedge\omega_{12}=\omega^*_{12}\wedge\omega_{12}.
\end{align*}
The curvature two-forms, or tensors, are determined by the matrix elements of the group.  Curvature is equivalent to force -- bosons are the physical carriers of force -- therefore, the matrix elements represent bosons.  This is consistent with the initial assertion that the group conveys interactions between subspaces of the flag.  While calculated here for a Grassmannian, this will be shown to hold for flag manifolds in general.

The metric and curvature are easily specialized to the $Sp(2)/Sp(1)^2$ case.  The curvature forms are particularly interesting.   In this simple case, define the scalar-vector one-form as $\omega_{12}=\omega=w_0{\bf 1}+{\bf w}$, so that 
\begin{align}
\label{Omegas}
\begin{split}
&\Omega_1=\omega\wedge\bar\omega=-2w_0\wedge {\bf w}-{\bf w}\wedge {\bf w}\\
&\Omega_2=\bar\omega\wedge\omega=+2w_0\wedge{\bf w}-{\bf w}\wedge {\bf w}
\end{split}
\end{align}
which are anti-self-dual and self-dual two-forms, respectively, as is proved in Appendix 1. It makes no difference to the physics which is which. (The reader may want to map this into Atiyah's representation, $q\sim X$, with use of the relations developed in the section on the metric.)  The curvature forms sit on the diagonal, which means that they are associated with the individual components of $V_2(\mathbb H)$. If one identifies $w_0$ with a  time-like quantity (which makes an analogy with special relativity), the interaction between the two particles (Alice and Bob) moves forward as seen by Alice and backward as seen by Bob.  Having entertained this thought, it is promptly dropped; there is a global time parameter that enters the picture, as claimed in ``Structural Preliminaries".  Regardless of interpretation, the splitting of interactions into self-dual and anti-self-dual partners is a general phenomenon and will recur for three particles.
 
\subsection*{Lie Algebra and Infinitesmal Generators}
A general matrix $x_q\in\mathfrak x$ in the Lie algebra $\mathfrak x$ of the coset $Sp(2)/Sp(1)^2$ is 
\begin{equation}\label{s2}
x_q=\left[{\begin{array}{*{20}c}
0& q \\
-\bar q &0 \\
\end{array}}\right],
\end{equation}
where $q$ is an arbitrary quaternion. The commutator $[x_a,x_b]$ is 
\begin{equation}\label{comm}
[x_a,x_b]= \left[{\begin{array}{*{20}c}
-[a,b^*]^*&0 \\
0 &-[a^*, b]^* \\
\end{array}}\right]
\end{equation}
where $[a, b^*]^*=ab^*-ba^*=ab^*-(ab^*)^*$ might be called a conjugating commutator; it is just the vector (imaginary) part of the $ab^*$ product: $[a, b^*]^*=2\Im(ab^*)$. In scalar-vector notation the diagonal elements are 
\begin{equation}\label{comm}
-[a,b^*]^*=2(a_0{\bf b}-b_0{\bf a}+{\bf a\times b}),\quad -[a^*,b]^*=2(-a_0{\bf b}+b_0{\bf a}+{\bf a\times b})
\end{equation}
and note the sign difference in the terms containing the identity components of the respective quaternions, reflecting those of the curvature two-forms.

The infinitesimal generators of the algebra are known for Grassmannians,\cite{BEE1} and displaying their commutators will  unite the matrix and operator descriptions.  Using the $SU(2)$ basis, the matrix representation of the quaternion $a$ appearing in eq. (\ref{conform}) is:  
\begin{equation}\label{conv1}
a=\left[{\begin{array}{*{20}c}
a_{11}&a_{12}\\
a_{21} &a_{22}\\
\end{array}}\right]=\left[{\begin{array}{*{20}c}
\zeta_1& \zeta_2\\
-\bar \zeta_2 & \bar \zeta_1\\
\end{array}}\right].
\end{equation}
The most convenient definition of the differential operator in the matrix representation is 
\begin{equation}\label{conv2}
\partial\sim[\partial_{\alpha a}]=\left[{\begin{array}{*{20}c}
\partial/\partial a_{11}&\partial/\partial a_{12}\\
\partial/\partial a_{21} &\partial/\partial a_{22}\\
\end{array}}\right]
\end{equation}
such that $\partial_{\alpha a}a_{\beta b}=\delta_{\alpha\beta}\delta_{ab}$.  Using different fonts for row and column indices is helpful for keeping track of terms when computing derivatives in multi-dimensional cases. 

The Lie algebra of the Grassmannian $Sp(n+k)/Sp(k)\times Sp(n)$ rendered in the $SU(2)$-basis might be denoted as $\mathfrak{cf}(k,n,\mathbb C^2)$.  It is parameterized by the elements $a_{\alpha b}$ of the $2k\times 2n$ Grassmannian matrix, and its generators are \cite{BEE1}: 
\begin{align}
h_{\alpha \beta}&= \Sigma_b [a_{\alpha b}\partial_{\beta b}-(a_{\alpha b}\partial_{\beta b})^*],\label{h},\;h_{\alpha \beta}\in \mathfrak{h=sp}(k);\\ 
H_{ab} & = \Sigma_\mu [a_{\mu a}\partial_{\mu b}-(a_{\mu a}\partial_{\mu b})^*],\;H_{ab}\in \mathfrak{H=sp}(n)\label{H},\\
p_{\alpha a}&=   \bar\partial_{\alpha a}+ \Sigma_{\mu b}a_{\alpha b}a_{\mu a}\partial_{\mu b};\label{p}
\end{align}
the summation convention is not used in this paragraph.  Note the remarkable fact that the generators within a subspace, the $\mathfrak h$ and $\mathfrak H$ components, are homogeneous operators, whereas those acting between subspaces, the $\mathfrak p$ components, are inhomogeneous.  This will be a key feature of the theory as it develops.  For completeness in this presentation, and to make a significant point later, the commutators are
\begin{align}
[(hJ)_{\alpha \beta},(hJ)_{\mu \nu}]={}&-J_{\alpha\mu}(hJ)_{\beta\nu}-J_{\alpha\nu}(hJ)_{\beta\mu}-J_{\beta\mu}(hJ)_{\alpha\nu}-J_{\beta\nu}(hJ)_{\alpha\mu}\notag\\
[(HJ)_{ab},(HJ)_{cd}]={}&-J_{ac}(HJ)_{bd}-J_{ad}(HJ)_{bc}-J_{bc}(HJ)_{ad}-J_{bd}(HJ)_{ac}\notag\\
[h_{\alpha \beta},H_{ab}]={}&0\notag\\
[(hJ)_{\mu\nu},p_{\alpha a}]={}&J_{\alpha\mu}p_{\nu a}+J_{\alpha \nu }p_{\mu a}\notag\\
[(HJ)_{bc},p_{\alpha a}]={}&J_{ab}p_{\alpha c}+J_{ac}p_{\alpha b}\notag\\
[p_{\alpha a},p_{\beta b}]={}&-J_{ab}(hJ)_{\alpha \beta}-J_{\alpha \beta }(HJ)_{ab}\notag\\
[\bar p_{\alpha a},p_{\beta b}]={}&\delta_{\alpha \beta}H_{ba}+\delta_{ab}h_{\beta \alpha}\label{pp}
\end{align}
The $J$-matrix factors,  see eq. (\ref{Jop}),  that are sprinkled throughout these equations make the symmetry of the commutation relations more apparent than they would be otherwise.  (The $2k\times 2k\; J$-factors with Greek indices are the tensor product of the $k$-dimensional identity with the ${\bf i}$ unit: $J_{\mu\nu}\sim{\bf 1}_k\otimes{\bf i}$.  $J$-factors with Roman indices are similar with dimension $2n\times 2n$.)  These equations are easily specialized to the case at hand: $n=k=1$.  

For our present concern, the Grassmannian consists of a single quaternion, which enables a considerable simplification.  Define the operators $\eta_j=\zeta_j\partial/\partial\zeta_j-\bar\zeta_j\partial/\partial\bar\zeta_j=-i\partial/\partial\theta_j$ for $\zeta_j=r_j\exp(i\theta_j)$: the components of the Cartan algebra from eqs. (\ref{h}) and (\ref{H}) are 
\begin{align*}
h_{11}=&\eta_1+\eta_2\\
h_{22}=&-h_{11}=-\eta_1-\eta_2\\
H_{11}=&\eta_1-\eta_2\\
H_{22}=&-H_{11}=-\eta_1+\eta_2.
\end{align*}
Substituting $\phi_\pm=(1/2)(\theta_1\pm\theta_2)$, and reverting to the quaternion basis, these are written succintly as 
\[
h_C=-{\bf k}\partial/\partial\phi_+\quad\textrm {and}\quad H_C=-{\bf k}\partial/\partial\phi_-,
\]
where the subscript $C$ is a reminder that these are in the Cartan algebra of the $\mathfrak{sp}(1)$ generators.  The eigenvectors of these operators are of the form $\exp(m{\bf k}\phi_\pm)$.  The point here is to show that the $\mathfrak {cf}(2)$ algebra contains two intertwined copies of the $\mathfrak{su}(2)$ algebra.  Cartan algebras are critical to the analysis of representations of $CF(n)$ just as they are for any group.  

\subsection*{Eigenvalues and Eigenvectors}
It will be beneficial to start with this simple case to develop some algebraic tools.  Select a $g\in Sp(2)$ and a corresponding $x\in \mathfrak {x\in g \backslash h}\sim\mathfrak {cf}(2)$ with an explicit representation as in eq. (\ref{s2}).  To reinforce the previous section, an element $g$ in the fundamental representation of the group is parameterized by $g=\exp(x)h$, where 
\[
h=\left[
\begin{array}{ccc}
 h_1&0\\
0&h_2\\
\end{array}
\right]
\]
with $h_i\in Sp(1), 1\le i\le 2$.  

Using eq. (\ref {s2}) for $x$, the diagonalization problem, $ xr=r\lambda, r\subset \tau$, yields two simple equations
\begin{align*}
-r_1\lambda + qr_2=&0\\
-\bar qr_1-r_2\lambda = &0.
\end{align*}
Multiply the first by $\bar q$ on the left and the second by $-\lambda$ on the right and add to get $r_2(|q|^2+\lambda^2)=0$, which follows because $|q|$ is a multiple of the identity and commutes with a quaternion. The non-trivial solution of this equation is $\lambda=\pm{\bf e}_m |q|$, where ${\bf e}_m$ is any one of the $\{{\bf i,j,k}\}$ basis elements.  In conformity with the usual convention in physics, where the quaternion basis element ${\bf k}$ is associated with the diagonal ($z$-direction) in the $SU(2)$ representation, the choice $\lambda = \pm {\bf k}|q|$ is made.  Normalizing the eigenvector gives $r_2=\pm q^{-1}r_1{\bf k}|q|$.  But $q^{-1}=\bar q/|q|^2$, so that $r_2=\pm\bar ur_1{\bf k}$, where $u=q/|q|$ is a unit quaternion. 

It is clear that the norm $|q|$ factors from the eigenvalue/eigenvector equation, so that one is left with just the orthonormalization condition for the eigenvalues.  A bit of experimentation with the algebra of this problem leads to a simpler structure if one writes $x$ as 
\[
x=|q|\left[
\begin{array}{ccc}
 0&{\bf k}\bar u\\
 u{\bf k} &0\\
\end{array}
\right]
\]
Using this representation of the coset it quickly follows that the matrix 
\[
\tau=\frac{1}{\sqrt{2}}\left[
\begin{array}{ccc}
 {\bf 1}&{\bf 1}\\
 u &-u\\
\end{array}
\right]\in Sp(2),
\]
comprises the eigenvectors of $x$ with eigenvalues $\lambda_\pm=\pm|q|{\bf k}$.  (Since these eigenvalues are valid for all $x\in\mathfrak x$, it is somewhat superfluous to instantiate with``select an $x\in\mathfrak x$", so this distinction between an element of the algebra and the entire algebra will be dropped in subseqent equations.)  The trick of representing the elements in the algebra in a particular way so as to facilitate the construction of the eigenvectors will also be used for the $CF(3)$ case, where it will be seen to have deep physical significance.  

The fact that one of the components of the vector space is trivial (the commutative basis element $\bf 1$) may explain the apparent mismatch in degrees of freedom between the formulation of the Yang-Mills theory for a single isospin and the $Sp(2)$ context in which it resides.  The identity component of $\tau$ is trivial, and is effectively submerged in the Yang-Mills formulation.  The visible (isotopic) spin content is in the $u$ factors.

The most natural physical theory evolves the group along a geodesic with time $t$, which recommends that one introduce a frequency $\omega$ to write $\omega t=|q|$, so that
\begin{equation}\label{Sp(2)}
X=\exp(t\mathfrak x)=\left[
\begin{array}{ccc}
 \cos(\omega t){\bf 1}&\sin(\omega t){\bf k}\bar u\\
 \sin(\omega t)u{\bf k}& \cos(\omega t){\bf 1}\\
\end{array}
\right]=\left[
\begin{array}{ccc}
 \cos(\omega t){\bf 1}&\sin(\omega t)v\\
- \sin(\omega t)\bar v& \cos(\omega t){\bf 1}\\
\end{array}
\right].
\end{equation}
The presence of the unit (real) component in the eigenvectors is very interesting.   An $m$-fold tensor product of $X\in Sp(2)/Sp(1)^2$, eq. (\ref{Sp(2)}), generates terms in ascending powers: ${\bf 1}, v, v^2,\cdots, v^m$ and their conjugates, together with trigonometric phase factors. These might be tentatively identified as primitive lepton states, for example, $v$ for the electron, $v^2$ for muon, and $v^3$ for the $\tau$ meson, . . ., with the conjugates being states of opposite chirality.  However, these simple-minded assignments are probably not correct, as there is an important additional fact that has to be introduced, and which will make the physics much more interesting.

\subsection*{Centralizer}
The centralizer of the torus consists of matrices that commute with the matrix of eigenvalues of the group; in this case the centralizer $C$ consists of matrices of the form 
\[
C_a=\left[
\begin{array}{ccc}
 \nu_1&0\\
 0 &\nu_2
\end{array}
\right]\quad \textrm{or}\quad C_b=\left[
\begin{array}{ccc}
 0&\nu_3\\
 \nu_4 &0
\end{array}
\right]
\]
with $\nu_{\alpha}=\cos(\varphi_{\alpha}){\bf 1}+\sin(\varphi_{\alpha}){\bf k}, \alpha=1,2$ or $ \nu_{\alpha}=\cos(\varphi_{\alpha}){\bf i}+\sin(\varphi_{\alpha}){\bf j}, \alpha=3,4$.  The centralizer exists for all $Sp(n)$ and will likely have a special place in particle theory.  

To see what role the centralizer might play, and to build more tools to apply to assignments, requires further development of the theory, so rather than continue with $CF(2)$ we turn attention to the more interesting three body problem to reveal yet more structure. In any case, a rigorous construction of representations will be a major but rewarding undertaking, and is left to the experts.

\section*{Case 2: $Sp(3)/Sp(1)^3$}
Given the success of $SU(3)$ in organizing meson and baryon states, the hope is that $CF(3)$, which includes spin degrees of freedom and apparently contains an $SU(3)$ subgroup, will provide additional insight into the structure of these composite particles.  This will be now be demonstrated.

\subsection*{Metric and Curvature}
This case is the simplest example of a flag manifold that is not also a Grassmannian.  To make the geometrical part of the presentation general, the metric and curvature for an arbitrary flag manifold will be presented, with specialization to $Sp(3)/Sp(1)^3$ left to the end.

Define a partition $\{k_1,k_2, \cdots, k_m\}$ of $n$ such that $\sum_\mu k_\mu=n$, and consider the flag manifold $Sp(n)/Sp(k_1)\times Sp(k_2)\times\cdots\times Sp(k_m)=Sp(n)/\bigotimes_\mu Sp(k_\mu)$.  Corresponding to this partition, the left invariant one-form $\omega$ is partitioned into block form
\[
\omega=g^*dg=\left[
\begin{array}{cccc}
 \omega_{11}&\omega_{12}&\cdots&\omega_{1m}\\
 \omega_{21}&\omega_{22}&\cdots&\omega_{2m}\\
 \vdots&\vdots&\ddots&\vdots\\
  \omega_{m1}&\omega_{m2}&\cdots&\omega_{mm}
\end{array}
\right]
\]
where $\omega_{\mu\nu}$ is a $k_\mu\times k_\nu$ block.  Just as for the Grassmannian, the scalar metric on the flag manifold is constructed from the squares of the blocks in the upper triangle as 
\[
ds^2=\sum_{1\le\mu\le\nu\le m}\textrm{tr}(\omega_{\mu\nu}\omega^*_{\mu\nu}).
\]

The curvature two-forms are again computed from the Maurer-Cartan equation just as was done for the Grassmannian, to give the curvature two-forms on the diagonal blocks
\[
\Omega_\alpha=-\sum^m_{\mu\ne\alpha} \omega_{\alpha\mu}\wedge\omega_{\mu \alpha},
\]
which generalizes Chern's calculation for the Grassmannian.\cite{Chern46}  The elements of  the lower triangle in $\omega$ are the negative conjugates of those in the upper triangle, so that 
\begin{equation}\label{curv}
\Omega_\mu=\sum_{\alpha<\mu} \omega_{\alpha\mu}\wedge\omega^*_{\alpha\mu }+\sum_{\alpha>\mu} \omega^*_{\alpha\mu}\wedge\omega_{\alpha\mu }.
\end{equation}
This signifies that subspace or system $V_\mu: v\in V_\mu= \{v\in V_{i+1}|v \notin V_i\}$ interacts with all other subspaces in the flag.  If the interaction of a single system with its surroundings is of interest, the components of the flag can be permuted so that the flag manifold reduces to a Grassmannian.

For the $Sp(3)/Sp(1)^3$ case at hand the symmetry of these equations is best displayed by making the change of notation: $\omega_{12}=\omega_c, \omega_{13}=-\bar\omega_b,\omega_{23}=\omega_a$, and labeling the curvature two-forms with corresponding Greek letters to write
\begin{align}
\Omega_\alpha=&\bar\omega_b\wedge\omega_b+\omega_c\wedge\bar\omega_c\nonumber\\
\Omega_\beta=&\bar\omega_c\wedge\omega_c+\omega_a\wedge\bar\omega_a\label{Omegas}\\
\Omega_\gamma=&\bar\omega_a\wedge\omega_a+\omega_b\wedge\bar\omega_b\nonumber
\end{align}
This set of equations is the most important result of this paper.  With this assignment of symbols a beautiful symmetry is revealed; each particle sees the other two particles, one with identity-containing components running forward and the other running backward. (Diagramable as a digraph on a triangle.) The use of ``forward" and ``backward" is an arbitrary assignment of labels to the signs of these components of the curvature two-forms, as was discussed for $Sp(2)/Sp(1)^2$.  The curvature two-forms comprise a 12-dimensional object, and given the oscillatory nature of the fundamental rep, this has a superficial resemblance to string theory.
 
\subsection*{Generators}
A general matrix $\mathfrak{x}$ in the Lie algebra of the coset $Sp(3)/Sp(1)^3$ may be parameterized by
\begin{equation}\label{coset3}
\mathfrak{x}=\left[{\begin{array}{*{20}c}
0& c & -b^* \\
-c^* &0 & a\\
b & - a^* & 0
\end{array}}\right].
\end{equation}
where $\{a,b,c\}$ are three linearly independent quaternions, represented for present purposes in the $SU(2)$ basis.  In constructing generators, we will use the conventions in eqs. (\ref{conv1},\ref{conv2}).  

The infinitesimal operators have to follow the pattern established in eq. (\ref{h}) for everything to be consistent.  But since the $CF(3)$ representation is acting on $V_3$, we can use the analogy with $SO(3)$ acting on $\mathbb R^3$ to build the family of operators (using a somewhat inelegant $(xy)$ notation, but using the summation convention), 
\[
(xy)_{\alpha\beta}=x_{\alpha a}\partial/\partial y_{\beta a}-\bar y_{\beta a}\partial/\partial\bar x_{\alpha a}
\]
with two others related by cyclic permutations from the set $\{x,y,z\}$. These generators have a nice symmetry property, as revealed by 
\[
(\overline{xy})_{\alpha\beta}=\bar x_{\alpha a}\partial/\partial \bar y_{\beta a}- y_{\beta a}\partial/\partial x_{\alpha a}=-(yx)_{\beta\alpha}, 
\]
which is written succintly as $(yx)=-(xy)^*$.  

The first commutator to evaluate is 
\begin{align*}
[(xy)_{\alpha\beta},(\overline{xy})_{\mu\nu}]=&[x_{\alpha a}\partial/\partial y_{\beta a}-\bar y_{\beta a}\partial/\partial\bar x_{\alpha a}, \bar x_{\mu b}\partial/\partial\bar y_{\nu b}-y_{\nu b}\partial/\partial x_{\mu b}]\\
=&[y_{\nu b}\partial/\partial x_{\mu b},x_{\alpha a}\partial/\partial y_{\beta a}]+[\bar x_{\mu b}\partial/\partial\bar y_{\nu b},\bar y_{\beta a}\partial/\partial\bar x_{\alpha a}]\\
=&\delta_{\alpha\mu}(yy)_{\nu\beta}-\delta_{\beta\nu}(xx)_{\alpha\mu}
\end{align*}
where 
\[
(xx)_{\alpha\beta}=x_{\alpha a}\partial/\partial x_{\beta a}-\bar x_{\beta a}\partial/\partial\bar x_{\alpha a}
\]
in obvious extension of the notation.   So, $(xx),(yy)$ and $(zz)$ clearly belong to the diagonal blocks of the matrix of generators, and since $(rr)^*=-(rr)$, the diagonal elements are pure imaginary. Note that with our convention for labeling matrix elements, $\partial x_{\alpha b}/\partial x_{\beta a}=\delta_{\alpha\beta}\delta_{ab}$ and $\partial\bar x_{\alpha b}/\partial x_{\beta a}=J_{\alpha\beta}J_{ba}$, with the latter a result of the conjugation operation $\bar x_{\alpha b}=J'_{\alpha\gamma}x_{\gamma c}J_{cb}$ in eq. (\ref{Jop}).  These generators differ from those used by Wallach\cite{Wal}.    

Completing the list of non-trivial commutators (but not writing those obtained by cyclic permutations), it is not difficult to prove that 
\begin{align*}
[(xx)_{\alpha\beta},(xx)_{\mu\nu}]=&\delta_{\beta\mu}(xx)_{\alpha\nu}-\delta_{\alpha\nu}(xx)_{\mu\beta}-J_{\beta\nu}(xxJ)_{\alpha\nu}-J_{\alpha\mu}(Jxx)_{\beta\nu}\\
[(xy)_{\alpha\beta},(xy)_{\mu\nu}]=&J_{\beta\nu}(xxJ)_{\alpha\mu}-J_{\alpha\mu}(Jyy)_{\beta\nu}\\
[(xy)_{\alpha\beta},(yz)_{\mu\nu}]=&\delta_{\beta\mu}(xz)_{\alpha\nu}\\
[(xx)_{\alpha\beta},(xy)_{\mu\nu}]=&\delta_{\beta\mu}(xy)_{\alpha\nu}-J_{\alpha\mu}(Jxy)_{\beta\nu}
\end{align*}
Here $(Jxx)_{\alpha\beta}=J_{\alpha\gamma}(xx)_{\gamma\beta}$, and similarly for $(xxJ)$ and $(Jxy)$, are symmetrized versions of the operators.  While messy, the first commutator can be shown to satisfy the usual relations for $\mathfrak{su}(2)$. Together with conjugation, these are all the tools that are needed to construct the complete set of generators for the $\mathfrak{cf}(3)$ algebra.  The last commutator can be used to show that $(xy)_{\alpha\beta}$ is a root vector, as are the other two operators $(yz)_{\alpha\beta}$ and $(zx)_{\alpha\beta}$. However, the off-diagonal operators are not linearly independent because any two generate the third.  It can also be seen that the operators on the diagonal sum to zero.  This implies a deep relation between the rank-two root spaces of $CF(3)$ and $SU(3)$, which will be explored later.  

Given that the infinistesmal generators are homogeneous operators of degree zero, it follows that the irreducible representations of $Sp(3)$ will be constructed from polynomials of degree $m$ in the three parameters; 
\[
\Psi_{ijk}(x,y,z)=x^iy^jz^k\pm\textrm{perm};\;i+j+k=m,
\]
where the permutations are over the other orders of the factors.  Representations may include conjugates as well as basis elements, as will be seen.  Permutations are essential because there is no physical reason for preferring one order of factors over another within a given symmetry class.  The symmetries of the representations can be tracked with Young diagrams, but that is not pursued here.  We will return to consider the representations after solving for the eigenvalues.

\subsection*{Curvature Operators}
The metric on the flag manifold is bi-invariant, which enables a straightforward calculation of the curvature tensor on the tangent space, mirroring the Cartan-Chern calculation on the co-tangent space.  Given three left invariant vector fields $X,Y,Z$, the curvature endomorphism $R(X,Y)$ is given by $R(X,Y)Z=-(1/4)[[X,Y],Z]$.\cite{Sternberg}. The Lie algebra consists of left-invariant vector fields, allowing us to make the identifications $X=(yz),Y=(zx), Z=(xy)$.  Working out just one of the three operators, we have 
\begin{align*}
-4R(X,Y)Z\rightarrow&[[(yz)_{\alpha\beta},(zx)_{\mu\nu}],(xy)_{\rho\sigma}]\\
=&[\delta_{\beta\mu}(yx)_{\alpha\nu},(xy)_{\rho\sigma}]\\
=&-\delta_{\beta\mu}[(\overline{xy})_{\nu\alpha},(xy)_{\rho\sigma}]\\
=&\delta_{\beta\mu}[\delta_{\sigma\nu}(yy)_{\alpha\rho}-\delta_{\rho\alpha}(xx)_{\sigma\nu}]
\end{align*}
with use of the commutators above.  The important points to note are that these operators are on the diagonal where they act on the individual components of the flag, and that they each (including the other two operators obtained by cyclic permutation) consist of terms with positive and negative signs, just as was seen for the co-tangent space version.

\section*{Diagonalization of $Sp(3)/Sp(1)^3$}
A fixed element $\mathfrak x=\mathfrak x_3$ from the Lie algebra of $Sp(3)/Sp(1)^3$ is parameterized by a matrix of the form in eq. (\ref{coset3}).  The eigenvectors $t\subset \tau$ are the solutions of $\mathfrak{x}t - t\lambda =0$, giving 
\begin{eqnarray}\label{eigeneq}
\nonumber-t_{1}\lambda + ct_2 - \bar bt_3={}0 \\
-\bar ct_{1}-t_{2}\lambda + at_3={}0\\
\nonumber bt_{1} - \bar at_2- t_{3}\lambda={}0 
\end{eqnarray}
(There should be no confusion between time $t$ and the eigenvectors denoted by the same symbol with subscript.)  Multiply the first of these equations by $\bar c$ from the left and the second by $-\lambda$ on the right and add to eliminate $t_1$.  Similarly, multiply the first by $b$ from the left and the last by $\lambda$ on the right and add to again eliminate $t_1$. This gives two equations from which, say, $t_3$ can be eliminated.  Similar operations to eliminate $t_1$ and then $t_2$ gives three equations, written symmetrically as 

\begin{align*}
u_{1}(\lambda^3+\lambda L^2) +(cab - \overline {cab})u_1 = {}0\\
u_{2}(\lambda^3+\lambda L^2) +(abc - \overline {abc})u_2 = {}0\\
u_{3}(\lambda^3+\lambda L^2) +(bca - \overline {bca})u_3 = {}0
\end{align*}
Here $L^2 = |a|^2+|b|^2+|c|^2$ and unit quaternions, $u_i$, have been substituted for $t_i=|t_i|u_i$ since the norms of the $t_i$ cancel.  These equations are obtained using only multiplication, inversion of real quaternions, addition and subtraction -- no determinant was computed.  The presentation is belabored to convey the care that has been taken with the non-commutative algebra.  

The occurrence of three different versions of the characteristic polynomial is illustrative of the well-known fact that the determinant of a quaternion matrix is ill-defined -- we have three different polynomials corresponding to different calculations of the term occupying the position corresponding to the determinant in the analogous problem over $\mathbb R$ or $\mathbb C$.    

With our choice of basis for the eigenvalues of a quaternion matrix, $\lambda = \hat\lambda{\bf k}$, where $\hat\lambda\in\mathbb R$ is a scalar; substituting this expression in the equations gives
\[
u_{i}(-\hat\lambda^3 +L^{2}\hat\lambda){\bf k}+{\bf d}_iu_i=0;    1\le i \le 3
\]
with the purely imaginary ${\bf d}_i$ defined in the obvious way from the three equations.  Multiplying on the right with ${\bf k}\bar u_i$ gives 
\begin{equation}\label{eig}
(\hat\lambda^2 -L^{2})\hat\lambda{\bf 1} +{\bf d}_i u_i{\bf k}\bar u_i=0.
\end{equation}
It is straightforward to see that $y{\bf k}\bar y$ has a vanishing identity component, since  rotation of the ``vector" part of a quaternion, in this case ${\bf k}$, by conjugation with a unit quaternion $y$ does not generate an identity component.  Now, since $(\hat\lambda^2 -L^{2})\hat\lambda$ is a scalar, ${\bf d}_iu_{i}{\bf k}\bar u_i$ must also be scalar.  Since $u_i{\bf k}\bar u_i$ is just the vector part of a quaternion,  represent it as ${\bf u}^{(2)}_i$, such that the product ${\bf d}_iu_{i}{\bf k}\bar u_i =-{\bf d}_{i}\cdot{\bf u}^{(2)}_i+{\bf d}_{i}\times{\bf u}^{(2)}_i$.  This has to be a scalar, which forces ${\bf u}^{(2)}_i$ to be parallel to ${\bf d}_i$.  Furthermore, $u_{i}{\bf k}\bar u _i$ is a unit quaternion, so that ${\bf u}^{(2)}_i=\pm{\bf {d}}_{i}/|d_i|$ and ${\bf d}_iu_{i}{\bf k}\bar u_i=\pm|d_i|$.  

So, the $u_i$ are defined by $d_i$, which is the usual situation, but for the fact that this problem has three different ``effective" determinants.  Some thought to this puzzle leads to the idea that it can be turned around to define the elements of $\mathfrak x$ in terms of the components of the eigenvectors, similar to what was done for the $Sp(2)/Sp(1)^2$ problem.  

\subsection*{The Simplification}
Given the $Sp(2)$ example, in which the algebraic operations were simplified with a particular representation of the group algebra, a bit of experimentation with the $Sp(3)$ problem just uncovered leads to the realization that a more convenient parameterization of the flag manifold algebra is
\[ 
\mathfrak{x}=L\left[{\begin{array}{*{20}c}
v_1& 0 & 0 \\
0 &v_2 & 0\\
0 & 0 &v_3
\end{array}}\right]\left[{\begin{array}{*{20}c}
0& w_3{\bf k}& w_2{\bf k} \\
w_3{\bf k} &0 &w_1{\bf k}\\
w_2{\bf k} & w_1{\bf k}& 0
\end{array}}\right]\left[{\begin{array}{*{20}c}
\bar v_1& 0 & 0 \\
0 &\bar v_2 & 0\\
0 & 0 &\bar v_3
\end{array}}\right]=LV(W\otimes{\bf k})V^*.
\]
where $v_i, 1\le i\le 3$ are three linearly independent unit quaternions, and $w_i\in \mathbb R^+:\Sigma_iw^2_i=1$ because $L^2=|a|^2+|b|^2+|c|^2$. The 12 real variables $\{a,b,c\}$ in eq. (\ref{coset3}) have been replaced by another 12 linearly independent variables.  Nonetheless, $\mathfrak{x}^*=-\mathfrak{x}$ as is required by the orthogonality condition.  

Let $M$ denote the matrix representation of $\mathfrak x$ with elements $m_{ij}=Lw_kv_i{\bf k}\bar v_j$. Then $\bar m_{ij}=Lw_kv_j{\bf \bar k}\bar v_i=-Lw_kv_j{\bf  k}\bar v_i=-m_{ji}$.  So, it is legitmate to identify the matrix elements in eq. (\ref{coset3}), the $\{a,b,c\}$ parameters, with the nicely symmetric products
\begin{equation}\label{btof}
a=Lw_1v_2{\bf k}\bar v_3;\quad b=Lw_2v_3{\bf k}\bar v_1;\quad c=Lw_3v_1{\bf k}\bar v_2.
\end{equation}

The matrix that diagonalizes $\mathfrak x$ by $\tau^*\mathfrak x\tau =\lambda$ is of the form 
$\tau=V(R\otimes{\bf 1})$, where the elements of $R$ are scalars.  The eigenvalue problem is reduced to 
\[
L(R\otimes{\bf 1})'V^*V(W\otimes{\bf k})V^*V(R\otimes{\bf 1})=(R\otimes{\bf 1})'(LW\otimes{\bf k})(R\otimes{\bf 1})=\hat\lambda\otimes{\bf k}
\]
where now 
\begin{equation}\label{RealEigen}
LR'WR=\hat\lambda
\end{equation}
is a matrix problem over real variables, with $R\in SO(3)$.  Note that use has been made of $(a\otimes b)(c\otimes d)=ac\otimes bd$, which is permitted because $R$ is a matrix of scalars.  It is easy to show that 
\[
X=\exp{\mathfrak x}=V(R\otimes{\bf 1})[\cos(\hat\lambda){\bf 1}+\sin(\hat\lambda){\bf k}](R\otimes{\bf 1})'V^*.
\]

The eigenvalues of eq. (\ref{RealEigen}) require solutions of 
\[
|W-\eta 1|=\det \left[{\begin{array}{*{20}c}
-\eta& w_3& w_2 \\
w_3 &-\eta&w_1\\
w_2 & w_1& -\eta
\end{array}}\right]=0
\]
or 
\begin{equation}\label{cubic}
\eta^3-\eta -2x = 0
\end{equation}
where $\hat \lambda=L\eta$ and $x=w_1w_2w_3$.  This simple equation captures all of eq. (\ref{eig}).  This parameterization of $CF(3)$ might be construed as the origin of the three color variables of chromodynamics: $\{w_1,w_2,w_3\}$.

\subsection*{Eigenvalues and Eigenvectors}

The discriminant of eq. (\ref{cubic}) is $\Delta=4(1-27x^2)$.  Given that $\Sigma_iw^2_i=1$, the parameterization $[w_1, w_2,w_3]=[\cos\alpha,\sin\alpha\cos\beta,\sin\alpha\sin\beta]$, enables the evaluation   $x=(1/4)\sin2\alpha\sin\alpha\sin2\beta$.  Since $Lw_i$ is a non-negative definite norm of the corresponding quaternion element of $\mathfrak x$, the angles $\alpha$ and $\beta$ are restricted to the first quadrant: $0<\alpha,\beta\le\pi/2$. If the discriminant $\Delta\ge 0$ there are three real roots, with multiple roots for the equality.  The maximum value that $x$ attains is at $\sin2\hat\beta=1$, and the extrema of $\sin2\alpha\sin\alpha$ occur at $2\cos2\hat\alpha\sin\hat\alpha+\sin2\hat\alpha\cos\hat\alpha=\sin\hat\alpha(2-3\sin^2\hat\alpha)=0$.  The minimum is at $\sin\hat\alpha=0$ and at the maximum $\sin\hat\alpha=\sqrt{2/3}$.  At the maximum, $\hat x=1/3\sqrt{3}$, proving that $\Delta \ge 0$: the solutions of the cubic are real.   At the maximum, $w_i=1/\sqrt{3}, 1\le i\le 3$.  

At the minimum at least one of the $w_i=0$ so that $x=0$; in this case $\eta=\{0,\pm 1\}$, and the representation is reducible.  The interesting aspect of this phenomenon is that $A_1$ interacting with $A_2$ interacting with $A_3$ is not sufficient to hold three particles together -- the interaction between $A_1$ and $A_3$ is also required to sustain the three particle state. This is the principle of detailed balance that we saw in the curvature two-forms.

The solutions of eq.(\ref{cubic}) are obtained from the identity
\[
4\cos^3\theta-3\cos\theta-\cos3\theta = 0.
\]
Set $\eta=(2/\sqrt{3})\cos\theta$, to find that 
\[
\cos3\theta=3\sqrt{3}x\le 1
\]
where the inequality follows from $x\le 1/3\sqrt{3}$.  The solutions are $\eta_{k+1}=(2/\sqrt{3})\cos(\theta_0+ 2k\pi/3), 0\le k\le 2$, where $-\pi/6<\theta_0<\pi/6$.  (The determinant vanishes at $\theta_0=\pm\pi/6$, which is excluded by the argument above.)  The three solutions are alternatively written as
\begin{eqnarray*}
\eta_1=&(2/\sqrt{3})\cos\theta_0;\quad 1< \eta_1\le 2/\sqrt 3\\
\eta_2=&-(1/\sqrt{3})\cos\theta_0+\sin\theta_0;\quad -1/\sqrt 3<\eta_2<0\\
\eta_3=&-(1/\sqrt{3})\cos\theta_0-\sin\theta_0;\quad -1<\eta_3<-1/\sqrt 3
\end{eqnarray*}
In this form the range of $\theta_0$ may be restricted to $0\le\theta_0<\pi/6$, as $\theta_0<0$ simply switches $\eta_2\leftrightarrow\eta_3$.  At $\theta_0=0$ there are two equal roots, and this simpler case is an easy calculation of eigenvectors.

The degenerate case $x=w_1w_2w_3=1/3\sqrt{3}$ gives vanishing discriminant, and this is only possible if $w=w_i=1/\sqrt{3}, 1\le i\le 3$.  The eigenvalues are $\{2/\sqrt{3},-1/\sqrt{3},-1/\sqrt{3}\}$, and a small calculation gives $r_1+r_2+r_3=0$ for $\eta=-1/\sqrt{3}=-w$, , and $r_1=r_2=r_3=w$ for $\eta=2/\sqrt{3}=2w$.  The matrix of eigenvectors, is
\[
R=\left[\begin{array}{ccc}
\frac{1}{\sqrt{3}}&\frac{1}{\sqrt{3}}&\frac{1}{\sqrt{3}}\\
\frac{1}{\sqrt{3}}&-\gamma&\delta\\
\frac{1}{\sqrt{3}}&\delta&-\gamma\\
\end{array}\right]; \quad \gamma=\frac{1}{2}(1+1/\sqrt{3}), \delta=\frac{1}{2}(1-1/\sqrt{3})
\]
which can be multiplied on the right by any $SO(2)\subset SO(3)$ that commutes with the matrix of eigenvalues.

The general case is a more interesting calculation.  The eigenvectors for $\theta_0\ne0$ satisfy
\[
 \left[{\begin{array}{*{20}c}
-\eta& w_3& w_2 \\
w_3 &-\eta&w_1\\
w_2 & w_1& -\eta
\end{array}}\right]\left[{\begin{array}{*{20}c}
r_1 \\
r_2\\
r_3
\end{array}}\right]=0
\]
with indefinite values for the $w_i$.  Calculations are facilitated by several useful identities, which can be derived from powers of  $[\textrm{Tr}(\eta)]=0$ and $\textrm{Tr}(\eta^k)=\textrm{Tr}(W^k)$. In addition to $\eta_1\eta_2\eta_3=2x$, the following are used extensively in the calculation of eigenvectors:
\begin{equation}\label{idents}
\textrm{Tr}(\eta^2)=2;\;
\textrm{Tr}(\eta^3)=6x;\;
\textrm{Tr}(\eta^4)=2;\;\sum_{\alpha<\beta}\eta_\alpha\eta_\beta=-1;\;\sum_{\alpha<\beta}\eta^2_\alpha\eta^2_\beta=1
\end{equation}

In solving for the $r_i$ it is convenient to define two combinations of the parameters: $b_i=\eta w_i+w_jw_k, \{i,j,k\}$ cyclic and $c_i=\eta^2-w^2_i$.  It is easy to prove that $b^2_i=c_jc_k$ with use of $\eta^3=\eta+2x$.  A bit of algebra yields the relations $r^2_i/r^2_j=c_i/c_j$, so that normalization of the eigenvectors yields 
\[
\Sigma r^2_i=1=[1+c_2/c_3+c_1/c_3]r^2_3.
\]
This, together with obvious symmetry, yields the solutions
\begin{equation}\label{r2}
r^2_{i\mu}=\frac{\eta^2_\mu -w^2_i}{3\eta_\mu^2-1},\;1\le i,\mu\le 3
\end{equation}
and this is clearly column normalized since $\Sigma_i r_{i\mu}^2=1$.  Proof that these components comprise a fundamental representation of $SO(3)$ is completed in Appendix 2.

The signs of the square roots in eq. (\ref{r2}) may be inferred from the eigenvalues.  For $\eta_1$ all components of the eigenvectors are positive.  For the negative eigenvalues, $\eta_2$ and $\eta_3$, at least one component of the corresponding eigenvectors is negative.  It appears that the general case will benefit from application of computer algebra; for now it suffices to observe that some components of $\tau$ are negative.  It is more interesting to move on to explore the physical consequences of the relation between eigenvectors and the representations discussed above.

The time-dependent representation of the coset $X\in CF(3)$ is constructed as 
\begin{align}\label{BD}
X=&V(R\otimes{\bf1})[\exp (\omega t\eta{\bf k})](R'\otimes{\bf1})V^*\\
X=&V[\cos(\omega tW{\bf 1})+\sin(\omega tW{\bf k})]V^*.\nonumber
\end{align}

\section*{Composite States}
Representations of isolated systems require a caveat.  The fundamental idea of the flag is that systems are interrelated through the action of the flag manifold.  Any representation of $CF(k)$ that is constructed in isolation will only capture intrinsic properties of the system, whereas extrinsic properties that derive from relations between a system and its surroundings are resolved in the larger question of how a $V_k$-system is imbedded in $V_{k+n}$ under the action of $Sp(k+n)/Sp(k)\times Sp(n)$.  A few more comments on this aspect of the theory will be made later.  

The infinitesimal generators of $\mathfrak{cf}(3)$ that have been developed are homogeneous operators, and it has been claimed that the irreducible representations are homogeneous polynomials of degree $m$ in three quaternion variables $\{x,y,z\}$.  These will be constructed from terms of the form
\[
\Psi(m_1,m_2,m_3)=x^{m_1}y^{m_2}z^{m_3}\pm \textrm{perm}:\; 0\le m_i\le m; \;\Sigma^3_{i=1}m_i=m
\]
where the permutations enable one to construct asymmetric, symmetric, and skew-symmetric states as appropriate to the choice of terms.  As stated before, in quaternion products there is no physical reason for distinguishing between, say, $xy$ and $yx$.  In this two-body case there are only symmetric and skew-symmetric representations to consider.  Expanding on this observation, we will indicate how particle states might  be constructed with two and three components.

The relation between bosons and fermions that is conveyed by the matrix structure of the fundamental representation means that a boson is written as a linear combination of products of fermions with the maximal torus, and this is also true in higher dimensional representations.  The representation will evolve along geodesics as required by the parameterization in eq. (\ref{BD}).  To execute a comprehensive program of assignments using these facts will require considerably more effort than can be accomplished within the scope of the present work. Nonetheless, it will be useful to indicate some promising directions with at least a few preliminary assignments and observations.  

For displaying composite states it is beneficial to show how the quaternion components sort themselves out, and the scalar-vector notation is convenient for this. In anticipation of these products, some of which involve powers of a quaternion, an alternative is to use $x=|x|u, u\bar u={\bf 1}$, with
\begin{equation}\label{chi}
u=\cos(\chi){\bf 1}+\sin(\chi){\bf r}; \quad {\bf rr=-r\cdot r=-1}.  
\end{equation}
Here are two useful theorems:  

Theorem 1: Given the vector components $\{{\bf a,b,c}\}$ of three quaternions, 
\[
{\bf a}\times({\bf b}\times{\bf c})={\bf b}(\bf a\cdot\bf c)-{\bf c}(\bf a\cdot\bf b),
\]
which is just the standard triple product from vector calculus;

Theorem 2: Powers of a unit quaternion $u: u\bar u = {\bf 1}$ are given by 
\[
u^k=\cos(k\chi){\bf 1}+\sin(k\chi){\bf r},
\]
where ${\bf rr}=-1$, is easily proved using eq. (\ref{chi}) with recursion.  Note that $\bar u=u^{-1}=\cos(\chi){\bf 1}-\sin(\chi){\bf r}$.

Tensor products of two copies of $X$ in eq. (\ref{BD}) will generate three copies of $\bf 1$ from $v_i\bar v_i$, three $\bar v_iv_j, i<j$, and three conjugates of the latter.  In constructing product states, we can work in either the algebra or the group, but given that velocity, and hence dynamics, are functions on the tangent space of any manifold, it is more immediately appealing and algebraically simpler to work in the tangent space, even though, as stated previously, dynamics is not currently within reach.  The simplification implies we are working in a neighborhood of the identity of the group.  In the following the scalar magnitudes of the $\mathbb H$-valued functions will frequently not be of immediate concern, which is not to say that the magnitudes are not important.  By introducing a frequency-like variable $\omega$ to partner with the time-like variable $t$ in the description of geodesics, as in eq. (\ref{BD}), the \emph{global} magnitude of the algebra was parameterized.  The $\mathbb H$-algebra between units of the group algebra and the module will be our focus.  Before starting a discussion of potential meson and baryon states, it will be useful to return to the $CF(2)$ case to offer some insights or conjectures that set the stage for further discussion.  

\subsection*{Leptons}
The simplest assumption that can be made is to identify a single (un-normalized) quaternion as a electron, $u\sim e$.  Given this, the muon cannot be simply $u^2$ and the $\tau\sim u^3$, since no assignment of masses would make sense with this assignment, nor do the spin states look right.  The next thing to try for a muon is to give $u$ a \emph{twist} with a factor of $\bf k$.  This is appealing because the decay of $\mu^-$ is almost exclusively into $e^-\bar\nu_e\nu_\mu$\cite{rpp}.  If the components of the centralizer are identified as neutrinos, a ${\bf k}u\sim \mu$ state only needs another factor of $\pm{\bf k}$ to produce an $e^-$.  On the other hand, if we are to imagine this state in the context of a much larger $Sp(n)$, it could be produced by the action of a very particular boson on an electron: $(u{\bf k}\bar u)(u)\to u{\bf k}$, would be interpreted as an electroweak interaction of a boson with an electron to produce a muon. (Concurrent with this, the conjugate boson acts elsewhere on the flag, but this is a topic for the future.)  This is not intended to be an experimental method for producing muons; the simple product  is only meant to  illustrate an algebraic operation that executes a transformation from one state to another.  To make an assignment for the $\tau$ meson requires more careful attention to magnitudes, as will be discussed later. There is also the option of multiplication by the off-diagonal $C_b$ components of the centralizer, which expands the options for converting states into one another.  

\subsection*{Products of Two Quaternions}
To begin the development, select two quaternions from the three quaternions extracted from the infinitesimal generators, giving six possibilities, symmetric and skew-symmetric: 
\begin{align*}
\Psi_\pm(x,y)=&(1/2)(xy\pm yx)\\
\Psi_\pm(y,z)=&(1/2)(yz\pm zy)\\
\Psi_\pm(z,x)=&(1/2)(zx\pm xz)\\
\end{align*}
where the factor of $1/2$ is a simple normalization.  In addition to these, the conjugates of each of the three quaternions are available for composing states.  This gives $4\times 6$ potential states. 

In addition to these states, we are allowed to construct symmetric and skew-symmetric states with inserted factors of ${\bf i,j,k}$.  For example,
\[
\pi_{xy} = (1/2)(x{\bf k}y+\bar y{\bf k}\bar x)
\]
has odd parity, since $P:\pi_{xy}=(1/2)(\bar y{\bf \bar k}\bar x+x{\bf \bar k}y)=-\pi_{xy}$.  Further, if one choses $y=\bar x$, this state becomes $\pi_x=x{\bf k}\bar x$.  If we identify $-\pi_{xy}$ as an anti-$\pi$, then this combination of quaternions qualifies as its own anti-matter state.  Now, does this state belong to the group algebra or the module?  It has the right components to belong to the algebra, but it has the wrong symmetry under the $PIN$ operators.  A matrix element $m_{ij}=u_i{\bf k}\bar u_j$ of the algebra transforms as follows:
\begin{align*}
&P: m_{ij}\to -u_j{\bf k}\bar u_i=-m_{ji}\\
&I:m_{ij}\to u_j{\bf k}\bar u_i=m_{ji}\\
&N:m_{ij}\to -m_{ij}
\end{align*}
However, $P:\pi_{xy}\to -\pi_{xy}$, which is the same as $N:\pi_{xy}\to -\pi_{xy}$, but $I:\pi_{xy}\to \pi_{xy}$ is equivalent to the identity operator, so $\pi_{xy}$ does not belong to the group algebra.  This simple calculation appears to be useful for identifying and categorizing terms.  Such aids are essential, because things have become quite complex, with many combinations possible; states constructed from an $a,b$ pair may contain many parity combinations as well as ${\bf i,j,k}$ factors.  

In computing explicit terms of products using the scalar-vector notation it is not difficult to see that many different combinations of terms, particularly symmetric states, will contain identity components, \emph{i.e.}, terms with the basis element ${\bf 1}$.  Using $\{a,b\}$ to be any pair of the $\{x,y,z\}$ fermions, the skew-symmetric combinations $(1/2)(ab-ba)={\bf a\times b}$ and $(1/2)(a{\bf k}b+\bar b{\bf k}\bar a)$ do not contain a term in the identity basis element, but the latter is a function of the identity components $\{a_0,b_0\}$ of the basic quaternions.  The reason for making a distinction between these two cases is the following: Define a quaternion 
\[
A(a,w)=A_0(a,w){\bf 1}+A_1(a,w){\bf i}+A_2(a,w){\bf j}+A_3(a,w){\bf k}=A_0{\bf 1}+{\bf A}
\]
where $a$ is a quaternion and $w$ is any set of parameters not including $a$, and also define the differential operator (see $\mathfrak p$ above)
\begin{equation}\label{del}
\partial_a=\partial/\partial a_0{\bf 1}-(\partial/\partial a_1{\bf i}+\partial/\partial a_2{\bf j}+\partial/\partial a_3{\bf k})=\partial/\partial a_0{\bf 1}-\nabla
\end{equation}
This acts on $\bar A$ to give
\[
\partial_a \bar A=(\partial A_0/\partial a_0-\nabla\cdot {\bf A}){\bf 1}-(\nabla A_0+\partial{\bf A}/\partial a_0)+\nabla\times{\bf A}
\]
which looks like $\partial_a\bar A= \xi{\bf 1}+{\bf E}+{\bf B}$, where ${\bf E,B}$ are the electric and magnetic fields of Maxwell theory.  (Further pursuit of this with a Wick rotation of $a_0$ to $it$ to get Maxwell's equations does not appear to be fruitful.  The quadratic piece of $\mathfrak p$ is non-Euclidean, and may be omitted near the origin of the coset.)  Now, the $ab-ba$ state has neither an identity component nor does it contain $a_0$ and $b_0$, so  the ``apparent" ${\bf E}$ vanishes, as consistent with a neutral particle.  A calculation gives $\partial_a({\bf a\times b})=-\nabla_a\times({\bf a\times b})=2{\bf b}$, so that this state has a magnetic moment.  I do not know if this corresponds to any known meson state.  

The symmetric state of $\{a,b\}$, denoted $\textrm{Sym}^2(a,b)$, is 
\begin{align*}
\textrm{Sym}^2(a,b)=&{}(a_{0}b_{0}-{\bf a\cdot b}){\bf 1}+(a_{0}{\bf b}+b_{0}{\bf a})\\
\textrm{Sym}^2(a,a)=&{}(a^2_0-{\bf a\cdot a}){\bf 1}+2a_{0}{\bf a}=a^2=\cos(2\theta_a){\bf 1}+\sin(2\theta_a){\bf u_a}.
\end{align*}
Applying the operator defined in eq. (\ref{del}) to the first of these gives
\[
\partial_a\textrm{Sym}^2(a,b)=4b_0{\bf 1}+2{\bf b}
\]
which appears to have an electric field but no magnetic moment.  It is premature to attempt to quantize these functions prior to a more thorough analysis of the infinitesimal generators -- these elementary calculations are merely intended to show how the various operators can be used to construct functions that have a classical interpretation.  Note that for both the skew-symmetric and symmetric functions, $f_\pm(a,b)$, that $\partial_a\partial_a f_\pm(a,b)=0$, so that the deeper correspondence with Maxwell's equations is not trivial.

\subsection*{Products of Three Quaternions}

In constructing homogeneous products of the fundamental $\{x,y,z\}$ quaternions, their scalar norms will factor and will be ignored for the present.  The unit quaternions will at first be given their meaning as the components of eigenvectors developed in the eigenvalue/eigenvector section.
\subsection*{Symmetric Products}
Totally symmetric states of $k$ quaternions can be formalized in $\textrm{Sym}^k$, the symmetrized product of $v_\mu, 1\le \mu\le k$, with
\begin{equation*}
k\textrm{Sym}^k(v_1,v_2,v_3,\cdots,v_k)=\Sigma_{1}^{k}v_{i}\textrm{Sym}^{k-1}(v_1,v_2,\cdots,\hat{v}_i,\cdots,v_k),
\end{equation*}
where $\hat{v}_i$ signifies that the term is missing. The first term in the sequence is $\textrm{Sym}^1(v_1)=v_1$.  The first few symmetric states are:
\begin{align*}
\textrm{Sym}^2(v_1,v_2)=&(v_1v_2+v_2v_1)/2,\;\textrm{consistent with the previous section}\\
\textrm{Sym}^3(v_1,v_2,v_3)=&[v_1\textrm{Sym}^2(v_2,v_3)+v_2\textrm{Sym}^2(v_1,v_3)+v_3\textrm{Sym}^2(v_1,v_2)]/3\\
=&[v_1(v_2v_3+v_3v_2)+v_2(v_1v_3+v_3v_1)+v_3(v_1v_2+v_2v_1)]/6
\end{align*}
A few explicit calculations will serve to show how the structure of symmetric and skew-symmetric products are different from one another.  For this purpose the $\{a,b,c\}$ set (arbitrary labels, not bosons) is reclaimed to avoid multiple subscripts.  The following may be readily verified:
\begin{align*}
\textrm{Sym}^3(a,b,c)=&(a_0b_0c_0-a_0{\bf b\cdot c}-b_0{\bf a\cdot c}-c_0{\bf a\cdot b}){\bf 1}\\
&+(b_0c_0-{\bf b\cdot c}/3){\bf a}+(a_0c_0-{\bf a\cdot c}/3){\bf b}+(a_0b_0-{\bf a\cdot b}/3){\bf c})\\
s_3(a,a,b)=&{}(a^2_0b_0-b_0{\bf a\cdot a}-2a_0{\bf a\cdot b}){\bf 1}+(a^2_0-{\bf a\cdot a}/3){\bf b}+2(a_0b_0-{\bf a\cdot b}/3){\bf a})\\
a^3=&{}(a^3_0-3a_0{\bf a\cdot a}){\bf 1}+(3a^2_0-{\bf a\cdot a}){\bf a}=\cos(3\theta_a){\bf 1}+\sin(3\theta_a){\bf u_a}
\end{align*}
Note that the symmetric functions all have identity components that contain the identity components of their constituents as well as components from the vector parts.  The function $s_3(a,a,b)$ has no symmetry with respect to interchange of its components, but is included here rather than separating it out. States composed with quarks having alternative chirality are not written down here, but they will be of interest. 

 \subsection*{Skew-Symmetric Products}
 
 The normalized skew-symmetric product, $\wedge^k$, of $k$ functions, $v_i, 1\le i\le k$, is 
\begin{equation}\label{skew}
n\wedge^k(v_1,v_2,v_3,\cdots,v_k)=\Sigma_{1}^{n}(-1)^{i-1} v_{i}\wedge^{n-1}(v_1,v_2,\cdots,\hat v_i,\cdots,v_k)
\end{equation}
By convention, $\wedge^k(v_1) = v_1$, and the first term of the sum on the right has a positive sign when the variables in the function are in sequential or lexical order. The first few skew-symmetric functions are 
\begin{align*}
\wedge^2(v_1,v_2)=&{}(v_1v_2-v_2v_1)/2\\
\wedge^3(v_1,v_2,v_3)=&{}[v_1(v_2v_3-v_3v_2)-v_2(v_1v_3-v_3v_1)+v_3(v_1v_2-v_2v_1)]/6
\end{align*}
As before, revert to the $a,b,c$ notation to avoid multiple subscripts to get the simplified skew-symmetric states:
\begin{align*}
{}&\wedge^2(a,b)={\bf a}\times {\bf b},\;\textrm{as before},\\
{}&\wedge^3(a,b,c)=-[\mathbf{a}\cdot(\mathbf{b}\times\mathbf{c})]+(1/3)[a_{0}(\mathbf{b}\times\mathbf{c})+b_{0}(\mathbf{c}\times\mathbf{a})+c_{0}(\mathbf{a}\times\mathbf{b})],\\
{}&\wedge^4(a,b,c,d)\equiv 0.
\end{align*}
There is no skew-symmetric state of four quaternions!  The proof is contained in Appendix 3.  The skew-symmetric state of three quaternions is very special in that it is a ``complete" quaternion, whereas the skew-state for two quaternions lacks a scalar part and is ``incomplete."  

One might augment these states of three quaternions with additional ones that are twisted with inserted $\bf k$ factors.  If this factor is included, the states begin to take on the character of linear combinations of products of fermions and bosons, which might be precursors to decomposition products.  It is difficult to avoid speculating, as so many possibilities are open for consideration.

The assignment of states in the flag environment will likely not coincide with currently accepted $SU(3)$ assignments with $\{u,d,s\}$ quarks, simply because the vector spaces on which the $Sp(3)$ and $SU(3)$ groups act are different.  For example, I have a strong suspicion that protons and neutrons correspond to $\wedge^3(v_1,v_2,v_3)$ and $\textrm{Sym}^3(v_1,v_2,v_3)$, but this awaits further analysis.

\subsection*{Why Six Quaternions?}
The collision of two baryons, protons for example, entails a strong contact between and mixing of states with content $\psi(v_1,v_2,v_3)$ and $\psi(v_4,v_5,v_6)$.  (The flag is currently silent on the dynamics of the collision.)  We have to expand the basis to six dimensions, which is why there are six quarks.  This enables the construction of states such as $\psi(v_1,v_2,v_4)$ and other combinations so as to cover states that are currently represented as, say, $udc$.\cite{rpp}  The mixing of these states also requires extension of the scheme to $Sp(6)/Sp(1)^6$.

Given that two protons are substates of a larger vector space, there is a small but non-vanishing probability that a third proton can become involved in a three-body collision.  If that were to happen, a seventh and even up to ninth quark would become evident.  A three body collision can only occur in highly concentrated counter-rotating proton beams, or in a collision with or of nuclei.  

\subsection*{Excited States \emph{vs.} Baryon States}
The first excited state of, say, a symmetric two-particle state with ground state $ab+ba$, is composed as $a^2b+ab^2+b^2a+ba^2$, which is different from the three quaternion ground state $a^2b+aba+ba^2$.  However, once normalized, these two different states should be comparable in some respects; for example, they might have closely similar masses (assuming that masses are assigned to the $a,b$ states).  The search for these close coincidences is part of the larger program of state assignments.  Another example that previously appeared is $\tau\sim u^3$ as an excited state, which begs to be compared with a ground state of three identical quarks. Assigning a mass to a $q$ with the replacement $u\to q$ for the $\tau$, a more appealing assignment is $\tau\sim q\bar q q=|q|^2q$.   Assuming the $SU(3)$ classifications\cite{rpp} hold, there are a few $\Delta$ states, $uuu$ and $ddd$, with masses around 1600-1800 MeV, as well the $\Omega\sim sss$ (1672 MeV), that are similar to the mass of $\tau\sim |q|^2q$ (1777 MeV), so give or take a few hundred MeV, this assignment for the $\tau$ meson seems reasonable.  

\section*{Discussion}
The overarching goal of interaction theory is to incorporate the states found for small systems into ever larger flags, to build up descriptions of more complex states of matter.  Some of the principles that are required for these extentions have been developed in the curvature forms and the Grassmannian generators.  The basic ideas are easily summarized:  The action of the fundamental representation $g\in Sp(n)$ on the state space $V_n(\mathbb H)$ by $g: V_n\to \hat V_n$ is a linear transformation of the space.  Eigenvectors of this action are stationary states; one dimensional subspaces are quaternion valued fermions, which transform as fundamental representations of the $Sp(1)\sim SU(2)$ group.   Any vector in the fundamental representation can be constructed as a linear combination of  eigenvectors.  The two and three particle examples that have been developed above show how the flag and flag manifold can be used to construct composite states.  

The representation of quantum systems with linear combinations of product states over $\mathbb {R,C}$ are well-known.  Extension of the algebra to the quaternion ring enables product states to be formed with spin content, which cannot be done with Dirac spinors.  The fundamental idea of this work is that elementary particles, represented by quaternions, comprise all matter, and that both stable and unstable composite particles are quaternionic functions of the fundamental units.  

The generators of the Lie algebra, $\mathfrak{cf}(3)$, will yield ladders of states having weights that are separated by integer values.  Transitions in systems that increase or decrease by an integer quantum number are understood to result from absorption or emission of a boson.  The generators within a subgroup are homogeneous, whereas inhomogeneous operators couple disjoint subspaces with one another.  This may help to explain the different categories of forces in particle physics.  
  
The small system calculations that have been done here utilized tricks that might not extend to $n>3$, but the representation of the bosonic matrix elements as linear combinations of products of fermions will be true for any $CF(\{ k_\mu\})$, where $\{k_\mu\}, \Sigma_\mu k_\mu=n$, is the partition of $n$ introduced previously.  The relation between bosons and fermions for the $n=2$ case is unique; for $n=3$ we found bosons as \emph{twisted products} of two fermions, the twist being the $\bf k$-component from the maximal torus.  For larger $n$ the matrix elements of the flag manifold are linear combinations of binary twisted  products of eigenvectors, so the relation between bosons and fermions carries over, it just becomes more complicated by linear combinations.

The crucial relation between $CF(3)$ and $SU(3)$ is that their root spaces are the same -- the characteristic polynomials of the algebras, $|\mathfrak{x}-\lambda 1|=0$, are isomorphic.  Thus, the success of $SU(3)$ in organizing elementary particle properties is expected to transfer to the $CF(3)$ representation. However, it may happen that the detailed assignments differ from one another, simply because the $V_3(\mathbb H)$ vector space picture is not identical to that of the $\{u,d,s\}$ quarks.  The extension to six quarks and more has been addressed but not pursued.

This work is but a first step in a very large endeavor.  In addition to an immediate interest in determining the structure of composite physical states, there are geometrical aspects, \emph{e.g.}, evolution of curvature of flag manifolds under Ricci flow, that appear to have significant physical implications.\cite{Che}.  The complexity of nature is revealed in the many ways that the elementary pieces fit together, and there is sufficient mathematical structure to yield insight in the forces that nature exhibits:  Within a subspace the infinitesimal generators of the Lie algebra are different operators from those that act between subspaces.

The tools that have been developed should be sufficient to begin an analysis of physical applications.  The Lie algebra will enable explicit representations to be constructed, and the $PIN$ involution operators will provide a basis for cataloguing states by their symmetries and quantum numbers.  A thorough classification of states in comparison with established meson and baryon assignments will be a major undertaking and is left to experts.  

\section*{Acknowledgement}
This work benefitted from conversations with Profs. John Sullivan, Univ. Washington, and Nolan Wallach, UCSD.

\section*{Appendix 1: Self-Dual and Anti-Self Dual Curvature Two-Forms}
The two curvature two-forms, $\Omega_1=\omega\wedge\bar\omega$ and $\Omega_2=\bar\omega\wedge\omega$, are anti-self-dual and self-dual, respectively, as will be shown.  A $k$-form $\Omega$ in a $2k$-dimensional space is self-dual (anti-self-dual) if the Hodge dual $*\Omega=+\Omega (-\Omega)$. Define the one-form $\omega=w_0+\bf w$ in scalar-vector notation, so that
\begin{align*}
\Omega_1= &\omega\wedge\bar\omega=-2w_0\wedge{\bf w}-{\bf w}\wedge{\bf w}\\
\Omega_2= &\bar\omega\wedge\omega=+2w_0\wedge{\bf w}-{\bf w}\wedge{\bf w}
\end{align*}
with which it follows that  
\begin{align*}
\Omega_1\wedge\Omega_1=&+4w_0\wedge{\bf w}\wedge{\bf w}\wedge{\bf w}\\
\Omega_2\wedge\Omega_2=&-4w_0\wedge{\bf w}\wedge{\bf w}\wedge{\bf w}.
\end{align*}
Since ${\bf w}=w_1{\bf i}+w_2{\bf j}+w_3{\bf k}$, it is seen that the only terms that survive the triple exterior product, ${\bf w}\wedge{\bf w}\wedge{\bf w}$,  are those with $\bf ijk=-1$ in some order. Ordering the quaternion basis in serial order in the triple product, and counting the permutations of terms with their symmetries, both with respect to the exterior algebra and the quaternion algebra, one finds 
\begin{align*}
\Omega_1\wedge\Omega_1=&-24w_0\wedge w_1\wedge w_2\wedge w_3{\bf 1}\\
\Omega_2\wedge\Omega_2=&+24w_0\wedge w_1\wedge w_2\wedge w_3{\bf 1},
\end{align*}
proving that the curvature two-forms are anti-self-dual and self-dual, respectively.  The reader will find the same result with use of the $SU(2)$ basis representation.

The proof that $\Omega_\mu=d\omega_{\mu\mu}+\omega_{\mu\mu}\wedge\omega_{\mu\mu}$ is a tensor over $\mathbb H$, \emph{i.e.}, $\Omega_\mu\to h_{\mu}^*\Omega_\mu h_{\mu}$ (no sum) with a change of basis of $V_n$: ${\bf e}_n\to {\bf e}_nH$ where $H=\bigoplus_\mu h_\mu$ preserves the stability subgroup and hence the flag structure, is the same as for the complex case.\cite{Chern95}

\section*{Appendix 2: Proof of $R\in SO(3)$ in Eq. (\ref{RealEigen})}
The proof that the matrix $R=(r_{i\alpha})\in SO(3)$ for the general case $x\ne 1/3\sqrt{3}$ is completed here.  In the text the normalization in columns was obvious. Normalization in rows requires 
\[
\Sigma_\alpha r_{i\alpha}^2=\Sigma_\alpha \frac{\eta_\alpha^2-w^2_i}{3\eta_\alpha^2-1}=1,\;1\le i\le 3
\]
which is proved with use of the relations in eq. (\ref{idents}).  Two intermediates that arise in this calculation,  
\[
\prod_\alpha(3\eta^2_\alpha-1)=4(27x^2-1)\; \textrm {and}\;\sum_{\alpha<\beta}(3\eta^2_\alpha-1)(3\eta^2_\beta-1)=0,
\]
are easily proved.  To prove orthogonality of the rows, begin with
\begin{equation}\label{ria}
r^2_{i\alpha}r^2_{j\alpha}=\frac{c_{i\alpha}c_{j\alpha}}{(3\eta^2_\alpha-1)^2}=\frac{b^2_{k\alpha}}{(3\eta^2_\alpha-1)^2}
\end{equation}
where a Greek index has been added to $b_i$ and $c_i$ to associate each with the corresponding eigenvalue.  The choice of the positive sign in the square root of eq. (\ref{ria}) gives
\[
\Sigma_\alpha r_{i\alpha}r_{j\alpha}=\Sigma_\alpha \frac{b_{k\alpha}}{3\eta^2_\alpha-1}=0; k\ne\{i,j\}, i\ne j,
\]
which is proved with use of the relations established above.  

The calculation of the column sum, $\Sigma_i r_{i\alpha}r_{i\beta}; \alpha\ne \beta$ is done by squaring, so that 
\[
(\Sigma_i r_{i\alpha}r_{i\beta})^2=\Sigma_i r^2_{i\alpha}r^2_{i\beta}+2\Sigma_{i<j} r_{i\alpha}r_{i\beta}r_{j\alpha}r_{j\beta}
\]
The cross-terms simplify, as the numerator of $r_{i\alpha}r_{j\alpha}$ is $(c_{i\alpha}c_{j\alpha})^{1/2}=\pm b_{k\alpha}$.  All of the denominators are the same, and with choice of the positive square root this simplifies to the calculation of the numerator 
\begin{align*}
T_{\alpha\beta}=&(3\eta^2_\alpha-1)(3\eta^2_\beta-1)(\Sigma_i r_{i\alpha}r_{i\beta})^2=\Sigma_i (c_{i\alpha}c_{i\beta}+2b_{i\alpha} b_{i\beta})\\
=&\Sigma_i[\eta^2_\alpha\eta^2_\beta-w^2_i(\eta^2_\alpha+\eta^2_\beta)+w^4_i+2w^2_i\eta_\alpha\eta_\beta+2x(\eta_\alpha+\eta_\beta)+2x^2/w^2_i]\\
=&3\eta^2_\alpha\eta^2_\beta-(\eta^2_\alpha+\eta^2_\beta)+\Sigma_iw^4_i+2\eta_\alpha\eta_\beta+6x(\eta_\alpha+\eta_\beta)+2x^2\Sigma_i1/w^2_i]
\end{align*}
The two remaining sums in $T_{\alpha\beta}$ are very simple: 
\[
\Sigma_i(w^4_i+2x^2/w^2_i)=\Sigma_iw^4_i+2\Sigma_{i<j}w^2_iw^2_j=(\Sigma_iw^2_i)^2=1,
\]
which gives
\begin{align*}
T_{\alpha\beta}=&3\eta^2_\alpha\eta^2_\beta-(\eta^2_\alpha+\eta^2_\beta)+2\eta_\alpha\eta_\beta+6x(\eta_\alpha+\eta_\beta)+1,\\
=&12x^2/\eta^2_\gamma-(2-\eta^2_\gamma)+4x/\eta_\gamma-6x\eta_\gamma+1,\\
\eta^2_\gamma T_{\alpha\beta}=&12x^2+\eta_\gamma(\eta_\gamma +2x)+4x\eta_\gamma-6x(\eta_\gamma +2x)-\eta^2_\gamma= 0.
\end{align*}
This completes the proof of orthonormality of the matrix $(r_{i\alpha})$.

\section*{Appendix 3:Proof that $\sigma_4(a,b,c,d)=0$}

Simplification of eq. (\ref{skew}) gives 
\[
24\sigma_4(a,b,c,d)=\{[a,b],[c,d]\}-\{[a,c],[b,d]\}+\{[a,d],[b,c]\}
\]
where $\{x,y\}=xy+yx$ is the symmetrizer. Since $[a,b]=2{\bf a}\times{\bf b}$, it follows that 
\[
3\sigma_4(a,b,c,d)= -({\bf a}\times{\bf b})\cdot({\bf c}\times{\bf d})+({\bf a}\times{\bf c})\cdot({\bf b}\times{\bf d})-({\bf a}\times{\bf d})\cdot({\bf b}\times{\bf c}).
\]
The cross-products, \emph{e.g.}, $({\bf a}\times{\bf b})\times({\bf c}\times{\bf d})+({\bf c}\times{\bf d})\times({\bf a}\times{\bf b})$, vanish in the symmetrizers.  Writing out just the ${\bf i}\cdot {\bf i}$ component of the three terms gives 
\begin{align*}
\textrm{coeff}({\bf i}\cdot {\bf i})={}& -(a_2b_3-a_3b_2)(c_2d_3-c_3d_2)\\
+&(a_2c_3-a_3c_2)(b_2d_3-b_3d_2)-(a_2d_3-a_3d_2)(b_2c_3-b_3c_2)\\
={}&(-a_2c_2b_3d_3+a_2d_2b_3c_3+b_2c_2a_3d_3-b_2d_2a_3c_3\\
{}&+a_2b_2c_3d_3-a_2d_2b_3c_3-b_2c_2a_3d_3+c_2d_2a_3b_3\\
{}&-a_2b_2c_3d_3+a_2c_2b_3d_3+b_2d_2a_3c_3-c_2d_2a_3b_3)
\end{align*}
in which it is seen that all terms cancel in pairs.  The remaining terms vanish by symmetry.

A much simpler proof is this: $\mathbb R^3$ does not admit four orthogonal vectors.

\end{document}